\pdfoutput=1
\RequirePackage{ifpdf}
\ifpdf % We are running pdfTeX in pdf mode
\documentclass[pdftex]{sigma}
\else
\documentclass{sigma}
\fi

\usepackage{bm}

\numberwithin{equation}{section}

\newtheorem{Theorem}{Theorem}[section]
\newtheorem{Corollary}[Theorem]{Corollary}
\newtheorem{Lemma}[Theorem]{Lemma}

{\theoremstyle{definition}
\newtheorem{Definition}[Theorem]{Definition}

}

\begin{document}

\allowdisplaybreaks

\renewcommand{\PaperNumber}{037}

\FirstPageHeading

\ShortArticleName{Building Abelian Functions with Generalised Baker--Hirota Operators}

\ArticleName{Building Abelian Functions \\ with Generalised Baker--Hirota Operators}

\Author{Matthew ENGLAND~$^\dag$ and Chris ATHORNE~$^\ddag$}

\AuthorNameForHeading{M.~England and C.~Athorne}

\Address{$^\dag$~Department of Computer Science, University of Bath, Bath, BA2 7AY, UK}
\EmailD{\href{mailto:M.England@bath.ac.uk}{M.England@bath.ac.uk}}
\URLaddressD{\url{http://www.cs.bath.ac.uk/~me350/}}

\Address{$^\ddag$~School of Mathematics and Statistics, University of Glasgow, G12 8QQ, UK}
\EmailD{\href{mailto:Christopher.Athorne@glasgow.ac.uk}{Christopher.Athorne@glasgow.ac.uk}}
\URLaddressD{\url{www.gla.ac.uk/schools/mathematicsstatistics/staff/christopherathorne/}}

\ArticleDates{Received March 16, 2012, in f\/inal form June 18, 2012; Published online June 26, 2012}

\Abstract{We present a new systematic method to construct Abelian functions on Jacobian varieties of plane, algebraic curves.  The main tool used is a symmetric generalisation of the bilinear operator def\/ined in the work of Baker and Hirota.  We give explicit formulae for the multiple applications of the operators, use them to def\/ine inf\/inite sequences of Abelian functions of a prescribed pole structure and deduce the key properties of these functions.  We apply the theory on the two canonical curves of genus three, presenting new explicit examples of vector space bases of Abelian functions.  These reveal previously unseen similarities between the theories of functions associated to curves of the same genus.}


\Classification{14H40; 14H50; 14H70}

\section{Introduction} \label{SEC:Intro}

The overall aim of this paper is the presentation of a systematic
method of construction of Abelian functions on Jacobian varieties of
plane, algebraic curves.  Apart from its purely theo\-retical interest
this is important from the point of view of at least two
applications. Firstly, the relations between Abelian functions of
given pole order, that is, the expansion of given functions in
terms of the basis, are partial dif\/ferential equations of the kind
arising in integrable systems theory. In particular, members of the KP
hierarchy have been shown to arise in this way, for functions associated with hyperelliptic curves of
general genus and also for various non-hyperelliptic curves.
Consequently, the Abelian functions constitute
one class of solution to such partial dif\/ferential equations. This
connection is explicit even in the early work of Baker (see later
references). Secondly, Abelian functions with shifted argument
expanded in terms of the basis of functions in un-shifted arguments
give addition laws generalizing the classic addition law of points
on the cubic curve. This expresses the group structure of the
Jacobian. It is also important in the theory of discrete integrable
systems.

The need for a new systematic method of construction is motivated in Section~\ref{SUBSEC1.2} by the so called basis problem for Abelian functions.  The solution we present is based on our def\/inition of generalised Baker--Hirota operators, following the work of Baker on Abelian functions and the work of Hirota in soliton theory.  Hence we continue the introduction in Section~\ref{SUBSEC1.1} with a~discussion of the Hirota derivative and how it motivates the general operators which follow.

\subsection{Hirota derivatives and operators} \label{SUBSEC1.1}

The Hirota derivative is originally a device belonging to the direct
method of solution to soliton equations, \cite{Hirota}.  A simple
paradigm is provided by the Korteweg--de Vries equation,
\[
u_t+u_{xxx}+6uu_x=0.
\]
Substitution for $u(x,t)$ by $2(\ln F(x,t))_{xx}$ and integration
with respect to $x$ (neglecting arbitrary functions of $t$) yields
the bilinear form
\[
D_x\big(D_t+D_x^3\big)F\cdot F=0,
\]
where the Hirota derivative is def\/ined as
\[
D_x^nD_t^mF\cdot G=
\partial^n_{x'}\partial^m_{t'}F(x+x',t+t')G(x-x',t-t')\big|_{x'=0,t'=0}.
\]
Thus, for example,
\begin{gather*}
D_xF\cdot G  = F_xG-FG_x, \\
D_xD_tF\cdot G  = F_{xt}G-F_xG_t-F_tG_x+FG_{xt}
\end{gather*}
etc. The $N$-soliton solutions arise via an ansatz of the form
$F(x,t)=f(e^{\theta_1},e^{\theta_2},\ldots,e^{\theta_N})$ where the
$\theta_i$ are linear in $x$ and $t$.

Of course, the Hirota ``derivative'' is not actually a derivative
since it does not act on products but on pairs as signalled by the
$\cdot$ above.  However, if we consider a product on function pairs
of the kind $(f,g)(h,k)=(fh,gk)$ and def\/ine $\mathcal{D}_x$ etc. as
Hirota operators that map from function pairs to function pairs,
\[
\mathcal{D}_x(f,g) = (\partial_xf,g)-(f,\partial_xg),
\]
then we do have a Leibniz derivation formula:
\[
\mathcal{D}_x\left((f,g)(h,k)\right)=\mathcal{D}_x(f,g)(h,k)+(f,g)\mathcal{D}_x(h,k).
\]
Because further, this Hirota operator is bilinear over addition and
scalar multiplication, it is sensible to treat the function pairs as
tensor products and the product as the standard multiplication on
tensor products:
\[
(f\otimes g)(h\otimes k)=fh\otimes gk.
\]
The original Hirota derivative is then a composition of
$\mathcal{D}_x$ with a total symmetrization operator,~$S$, mapping
from the tensor to the dif\/ferential polynomial algebra:
\begin{gather*}
\begin{array}{@{}ccccc}
F \otimes F & \stackrel{\mathcal{D}_x}{\longrightarrow} & F \otimes F                   & \stackrel{S}{\longrightarrow} & F, \\
f \otimes g & \stackrel{\mathcal{D}_x}{\longmapsto} & f_x \otimes g
- f \otimes g_x & \stackrel{S}{\longmapsto} & f_xg-fg_x.
\end{array}
\end{gather*}
Here $F$ denotes a $\mathfrak{k}$-algebra of appropriate
$\mathfrak{k}$-valued functions, for some f\/ield $\mathfrak{k}$.

Such a generalisation has been the subject of previous work by one
of the authors, applied to invariant theory \cite{CA2006}, the
theory of Pad\'e approximants \cite{CA2005} and to generalised
$\wp$-functions~\cite{CA2003}.

It is also interesting to compare with algebraic notions of
derivations over tensor products of algebras introduced in~\cite{Matsumura}. It is not clear to the present authors whether
$\mathcal{D}$ is either an example or a generalization of this
approach but it suggests an avenue for further investigation

In each of these applications, there is an equivariance property of
the Hirota operators appropriate to the case. Suppose we have a
group, $\mathcal{G}$ acting on the space of independent variables
represented by the list $x$.  We assume there is an induced
representation of $\mathcal G$ on some set of functions of $x$:
\[
f\big(G^{-1}x\big)=f^G(x).
\]
In some applications $\mathcal G$ is a Lie group and we are
interested in a Lie algebra representation.  Then a Hirota operator
$\mathcal{D}$, (we drop the subscript), should have the desirable
property:
\[
\begin{array}{@{}ccc}
F \otimes F                & \stackrel{G}{\rightarrow} & F \otimes F                \\
\mbox{\tiny{$\mathcal{D}$}}\downarrow  &                           & \mbox{\tiny{$\mathcal{D}$}}\downarrow \\
F \otimes F                & \stackrel{G}{\rightarrow} & F \otimes F
\end{array}
\]
In the case of the Abelian functions associated to a curve, just
such an action is obtained.  Given a curve of genus $g$ we def\/ine
the associated Abelian functions as those meromorphic functions of
$g$ complex variables $\bm{u} \in \mathbb{C}^g$, which are periodic with
respect to the period lattice of the curve.  All such functions may
be expressed algebraically using the $\sigma$-function of the curve.
A translation across lattice points, $\bm{u} \mapsto
\bm{u}+\bm{\ell}$ induces a multiplicative factor:
$\sigma(\bm{u}+\bm{\ell})=h(\bm{u},\bm{\ell})\sigma(\bm{u}).$  The
generalised Hirota operators we introduce here will generate Abelian
functions out of tensor products of these $\sigma$-functions.

Our generalisation of the Hirota operator picks up from earlier
work.  In \cite{GRH94}, \emph{trilinear} Hirota derivatives are
presented in the context of certain integrable equations in
trilinear form.  They are characterised by a simple equivariance
property.  Further work can be found in \cite{H96, HGR95}.

In \cite{CA2006} a more algebraic, independent treatment is
developed for Hirota type operators acting on arbitrary tensor
products of f\/inite-dimensional $\mathfrak{sl}_2(\mathbb C)$-modules.
In the inf\/inite dimensional limit the $\mathfrak{sl}_2$ action
becomes a Heisenberg action and the Hirota operators become the
familiar Hirota derivatives (when acting on degree two tensor
products).  This interplay has also been exploited in a study of
dif\/ferential normal forms, \cite{Olver_Sanders}.

In the present paper we deal with arbitrary tensor products of a
dif\/ferential algebra over some f\/ield, $\mathfrak{k}$, which may as
well be $\mathbb C$ given the analytic context.  The dif\/ferential
algebra is generated by a Kleinian $\sigma$-function in variables
$u_1,\ldots,u_g$ and its partial derivatives.  There is a Heisenberg
action on the derivatives with respect to each individual $u_i$ and
an $\mathfrak{sl}_2(\mathbb C)$ action on the whole set of $u_i$.
For example the set of f\/irst order derivatives
$(\partial_{u_1}\sigma,\partial_{u_2}\sigma,\ldots,\partial_{u_g}\sigma)$
is a~basis for a $g$ dimensional
$\mathfrak{sl}_2(\mathbb{C})$-module.

\subsection{The basis problem for Abelian functions} \label{SUBSEC1.2}

The main motivation for the generalised Hirota operators presented
in this paper is to solve the so called \textit{basis problem} for
Abelian functions.  The simplest Abelian functions are the elliptic
functions, associated with elliptic curves, which of course have
genus one.  Elliptic functions have been the subject of much study
since their discovery and have been extensively used to enumerate
solutions of non-linear wave equations.  They occur in many physical
applications; traditionally the arc-length of the lemniscate and the
dynamics of spherical pendulums, \cite{MKandM99}, but also in
cryptography, \cite{Washington08}, and soliton solutions to the KdV
equation, \cite{Dodd82}.

Recent times have seen a revival of interest in the theory of their
generalisations.  These Abelian functions are also beginning to f\/ind
a wide range of applications. For example, they give further
solutions to the KdV equation along with solutions to other
integrable equations from the KP-hierarchy (see for example
\cite{bego08,bel97,MEe09}). They have also been used to describe
geodesic motions in certain space-time metrics, \cite{EHKKL}.

Much progress has been made through the realisation of Abelian
functions as generalisations of the Weierstrass $\wp$-function,
following the approach of Klein and Baker for hyperelliptic
functions.  In the genus $g$ case the original Weierstrass
$\wp$-function is generalised to a family of functions denoted
$\wp_{ij}(\bm{u})$ for $i,j\in(1,\dots,g)$.  It is the case that
$\wp_{ij}=\wp_{ji}$ and so there are $\tfrac{1}{2}g(g+1)$ functions
in total.  We have the integrability property that
\[
\frac{\partial}{\partial u_i} \wp_{jk} = \frac{\partial}{\partial
u_j}\wp_{ik}.
\]
So we can denote the derivatives of the functions using additional
indices and adopt the notation of writing the indices in ascending
numerical order.  The 3-index $\wp$-functions generalise the
derivative of the Weierstrass function, $\wp'$ and so on.

We may classify the Abelian functions according to their pole
structure.  The $m$-index $\wp$-functions will have poles of order
$m$ for example.  We can def\/ine a vector space of Abelian functions
with poles of at most a given order and so the question of building
bases for the vector spaces naturally arises.  The Riemann--Roch
theorem for Abelian varieties gives the dimension of the vector
space as $m^g$ where $m$ is the maximal pole order.  In the genus
one case each successive vector space will have dimension one
higher, and the new basis function can be played by the next
derivative of the $\wp$-function.  However, for $g>1$ there are not
enough $\wp$-functions to build the bases.  Indeed, as both $g$ and
$m$ increase so does the def\/icit of functions.  This is the basis
problem.

Solutions to the problem have been found in a number of specif\/ic
cases.  In genus 2 the introduction of one further function with
poles of order 3 is suf\/f\/icient to allow dif\/ferentiation to generate
all subsequent bases.  This new function is the dif\/ference
$\wp_{11}\wp_{22} - \wp_{12}^2$, in which the poles of order 4 in
each term cancel.  In \cite{MEeo11, ME11} the basis problem was
solved in various cases, by a group including one of the present
authors, through the introduction of analogous functions;
polynomials in $\wp$ with coef\/f\/icients chosen to cancel poles.
However for curves with $g>2$
hyperelliptic and $g>3$ non-hyperelliptic the bases cannot be f\/initely generated by
dif\/ferentiation and this approach does not generalise easily to give
an inf\/inite number of functions.

Solving the basis problem makes it easier to work with the functions
in a wide range of applications.  For example, in \cite{MEg09} a
reduction for the Benney equation was constructed using the Abelian
functions associated with a genus six curve and having a basis for
the functions limited the search for the explicit form of the
reduction.  The bases can also be used to derive more theoretical
results, such as the dif\/ferential equations satisf\/ied by the
functions.  These can be derived from a set which generalise the
classic dif\/ferential equation for the Weierstrass $\wp$-function,
\[
\big(\wp'(u)\big)^2 = 4\wp(u)^3 - g_2\wp(u) - g_3.
\]
The bases can also be used to construct various addition formulae for
the functions.  For example, there are generalisations of the
following well known addition formula for the Weierstrass functions,
\[
- \frac{\sigma(u+v)\sigma(u-v)}{\sigma(u)^2\sigma(v)^2} = \wp(u) -
\wp(v).
\]
Each generalisation has the same ratio of $\sigma$-functions
expressed as a polynomial in an appropriate basis of Abelian
functions.  For curves with an extra cyclic symmetry in their
coordinates there are other classes of addition formulae.  For
example, suppose the curve is invariant under $[\zeta^{j}]: (x,y)
\rightarrow (x, \zeta^j y)$ where $\zeta$ is some primitive $m$th
root of unity.  Then we can def\/ine a corresponding action on the
variables $\bm{u}$ and f\/ind an expression in an appropriate basis of
Abelian functions for
\[
\prod_{j=1}^{m} \frac{\sigma\left( \sum\limits_{i=1}^m
[\zeta^{i+j}]\bm{u}^{[i]} \right) }{\sigma( (\bm{u}^{[j]})^m ) },
\]
where $\bm{u}^{[1]}, \dots, \bm{u}^{[m]}$ are dif\/ferent sets of
variables.  The existence of such formulae helped motivate the new
Abelian functions constructed in Section \ref{SEC:Rf}.  For examples
of using bases to calculate dif\/ferential equations and addition
formulae see \cite{bego08,MEeo11,eemop07,ME11,MEe09}.

While the $\wp$-functions may be def\/ined algebraically using the
curve, an alternative def\/inition using the $\sigma$-function of the
curve can be easier to work with.  Recall that Weierstrass
introduced an auxiliary function, $\sigma(u)$, in his theory which
satisf\/ies
\[
\wp(u) = - \frac{d^2}{d u^2} \log \big[ \sigma(u) \big].
\]
The Weierstrass $\sigma$-function is usually def\/ined as an inf\/inite
product over the periods, but it can also be expressed using the
f\/irst Jacobi $\theta$-function, multiplied by a constant and
exponential factor.  This def\/inition generalises naturally to give
higher genus $\sigma$-functions as multivariate functions def\/ined
using the Riemann $\theta$-function  Once the generalised
$\sigma$-function is def\/ined we may then consider generalised
$\wp$-functions in analogy to the previous equation:
\[
\wp_{ij}(\bm{u}) = - \frac{\partial^2}{\partial u_i\partial u_j}
\log \big[ \sigma(\bm{u}) \big].
\]
Then, using the notation above, we see this is equivalent to the
following def\/inition in Hirota operators:
\[
\wp_{ij} = \left(\frac{-1}{2\sigma^2}\right) \, S \circ
\mathcal{D}_{i} \circ \mathcal{D}_{j} \left( \sigma \otimes \sigma
\right),
\]
where $\mathcal{D}_i$ indicates that the dif\/ferentiation in the
Hirota operator is with respect to $u_i$.  This led to the
def\/inition of so called $Q$-functions, where the operator is applied
not twice but an even number of times, generating further Abelian
functions with poles of order~2.  See
\cite{bego08,eemop07,MEhgt10, MEe09} for examples of how such
functions have been used to complete pole~2 bases.  In this paper we
provide the f\/irst proof that such functions are always Abelian.

These $Q$-functions led to the derivation of the generalised Hirota
operators which are
introduced in this paper.  They allowed for
the def\/inition of more general class of Abelian functions which we
call $\mathcal{R}$-functions.  Inf\/initely many such functions can be
def\/ined with a given pole order, making them useful for the general
basis problem.

The paper is divided into two parts.  The f\/irst deals with the
theory of the operators and the second the use of this
theory in the construction of Abelian functions.  While it is this
application which is of primary interest in this paper, it is
anticipated that the operators may have other uses, particularly in
the world of integrable systems, and so we present their theory
separately where possible. Section \ref{SEC:D} reformulates the
original Hirota theory into a form that more easily generalises, and
proves some explicit results for the functions obtained by repeated
application of the operators.  Section \ref{SEC:H} then def\/ines the
generalised operators and derives some of their properties.
Sections \ref{SEC:Qf} and \ref{SEC:Rf} apply the operator theory to
def\/ine inf\/inite classes of Abelian functions, f\/irst with poles of
order 2 and then poles of arbitrary order.  Finally Section~\ref{SEC:Ex} uses these functions to construct new bases of Abelian functions for the two canonical curves of genus three.

 \section[The Baker-Hirota operator]{The Baker--Hirota operator} \label{SEC:D}

We start by f\/ixing some notation.  Throughout the paper we are
dealing with functions of $g$ variables.  We drop the variables from
our notation unless required.  So for example,
\[
f = f(\bm{u}) = f(u_1,u_2, \dots, u_g).
\]
When working with Abelian functions, $g$ will be the genus of some
underlying algebraic curve.  Until we come to the work in Section~\ref{SEC:Ex}, none of the calculations in this document will depend
on the particular curve or genus.  In this section and the next the
reader may assume we are just working in a $\mathbb{C}$-algebra of entire
functions and their tensor products $\bigotimes_{\mathbb{C}}$.

Later we use subscripts on a function to indicate the application of
operators, but for now we shall use subscripts to indicate
dif\/ferentiation with respect to one of the variables.  For example,
\[
f_i = \frac{\partial}{\partial u_i} f(u_1,u_2, \dots, u_g).
\]
Superscripts in square brackets indicate a label, with other
superscripts indicating the object raised to a power as normal.

\begin{Definition} \label{def:D}
We let $\mathcal{D}_i$ denote the \textit{Baker--Hirota operator}.
This acts on the tensor product of two functions as
\begin{gather*} %\label{eq:D}
\mathcal{D}_i: \  f^{[1]} \otimes f^{[2]} \longrightarrow f^{[1]}_i
\otimes f^{[2]} - f^{[1]} \otimes f^{[2]}_i.
\end{gather*}
\end{Definition}
This operator is the key component in the famous Hirota derivative,
discussed in the introduction.  We refer to it as a Baker--Hirota
operator since it was f\/irst used by Baker in \cite{BA03}.

 We note that Baker--Hirota operators commute with each
other,
\[
\mathcal{D}_i \circ \mathcal{D}_j\big(f^{[1]} \otimes f^{[2]}\big) =
\mathcal{D}_j \circ \mathcal{D}_i\big(f^{[1]} \otimes f^{[2]}\big).
\]

\begin{Definition} \label{def:S}
We let $S$ denote the $\mathbb{C}$-linear \textit{symmetrization
operator} which will send a~tensor product to standard
multiplication,
\begin{gather*} %\label{eq:S}
S: \  f \otimes g \longrightarrow fg = gf.
\end{gather*}
\end{Definition}
We are usually interested in applying the Baker--Hirota operator to a
tensor product of the same object.  We note that the symmetrization
operator would annihilate the resulting sum after the application of
a single operator,
\[
S \circ \mathcal{D}_i(f \otimes f) = S\big( f_i \otimes f - f
\otimes f_i \big) = f_if - ff_i = 0.
\]
However, when two operators are applied, there is a non-trivial
result left after symmetrizing,
\begin{gather*}
S \circ \mathcal{D}_i \circ \mathcal{D}_j (f \otimes f)
= S\big( (f_{ij} \otimes f - f_i \otimes f_j) - (f_j \otimes f_i - f \otimes f_{ij}) \big) %\\&= f_{ij}f - f_if_j - f_jf_i + ff_{ij}
= 2(ff_{ij}-f_{i}f_{j}).
\end{gather*}
We will now consider the ef\/fect of multiple applications of a
Baker--Hirota operator. We use $I_n = \{ i_1, i_2, \dots , i_n \}$
for the set of indices of the operators, assumed distinct for the
moment.  We introduce the multiple sum notation,
\[
\sum_{1 \leq j_1<j_2<\dots<j_m \leq n} = \sum_{j_1=1}^n
\sum_{j_2=j_1+1}^n   \cdots   \sum_{j_m=j_{m-1}+1}^n.
\]
Note that if there are no values of an index $j$ in the appropriate
range then that sum does not run.

\begin{Lemma} \label{lem:Dntimes}
For $n \geq 1$ we have
\begin{gather} \label{eq:Dntimes}
\mathcal{D}_{i_1} \circ \cdots \circ \mathcal{D}_{i_n} \big( f
\otimes g \big) = \sum_{m=0}^{n} (-1)^m \!\left(
\sum_{1 \leq j_1<j_2<\dots<j_m \leq n} \!\! f_{I_n \backslash \{
i_{j_1},i_{j_2},\dots,i_{j_m} \} } \otimes g_{i_{j_1}i_{j_2}\dots
i_{j_m}} \right).\!\!\!
\end{gather}
When $m=0$ the entire inner sum is interpreted as a single term with
the set $\{ i_{j_1},i_{j_2},\dots,i_{j_m} \}$ being the empty set,
$\varnothing$.  A function with an empty set of subscripts is just the
function itself.
\end{Lemma}

 We give the details of an inductive proof in Appendix~\ref{APP_lemDntimes}.

The lemma allows us to observe the following result, detailing when
the symmetrization operator annihilates the result of Baker--Hirota
operators applied to the same function.

\begin{Corollary} \label{cor:Dnodd}
If $n$ is odd then $ S \circ \mathcal{D}_{i_1} \circ \cdots \circ
\mathcal{D}_{i_n} \big( f \otimes f \big) = 0$.
\end{Corollary}
\begin{proof}
We will show that when $n$ is odd, symmetrizing in equation
(\ref{eq:Dntimes}) causes all terms to cancel.  For example, the
f\/irst term, $f_{I_n} \otimes f$, and the last term, $(-1)^{n}f
\otimes f_{I_n}$ will cancel with each other.

First note that if $n$ is odd then the outer sum in
(\ref{eq:Dntimes}) will have an even number of terms.  We may
consider these terms pairwise, when $m$ is $m_1=\lfloor \frac{n}{2}
\rfloor -k$ and $m_2=\lceil \frac{n}{2} \rceil +k$, for some integer
$k$. So each pair consists of the sums
\[
(-1)^{m_1}   \sum_{1 \leq j_1 < j_2 < \dots  < j_m
\leq n}  f_{I_n \backslash \{
i_{j_1},i_{j_2},\dots,i_{j_{m_1}} \}} \otimes
f_{i_{j_1},i_{j_2},\dots, i_{j_{m_1}}}
\]
and
\[
(-1)^{m_2}   \sum_{1 \leq j_1 < j_2 < \dots < j_m
\leq n} f_{I_n \backslash \{
i_{j_1},i_{j_2},\dots,i_{j_{m_2}} \}} \otimes
f_{i_{j_1},i_{j_2},\dots, i_{j_{m_2}}}.
\]
Now, $n$ is odd so the f\/loor, $\lfloor \, \rfloor$ and ceiling
$\lceil \, \rceil$ of $\tfrac{n}{2}$ will be of opposite relative
parity, and hence so will $m_1$ and $m_2$ as they dif\/fer from the
f\/loor and ceiling by the same integer $k$.

Now since
\[
n = \left\lfloor \frac{n}{2} \right\rfloor + \left\lceil \frac{n}{2}
\right\rceil,
\]
we have
\[
n-m_1 = \left( \left\lfloor \frac{n}{2} \right\rfloor + \left\lceil
\frac{n}{2} \right\rceil \right) - \left( \left\lfloor \frac{n}{2}
\right\rfloor -k \right) = \left\lceil \frac{n}{2} \right\rceil + k
= m_2,
\]
and similarly $n-m_2=m_1$.  Hence the partition of indices on either side of the tensor in the two terms are complementary.  When we symmetrize they will give the same sum of terms, and so the dif\/fering signs outside the sums will cause them all to cancel.
\end{proof}

 Similar calculations will verify that the object is
non-zero when $n$ is even.

We now introduce a multiplication on tensor products, which simply
creates a tensor product of the same length by multiplying
corresponding entries.  For example,
\begin{gather*}
%\label{eq:cdotnot}
(a \otimes b)(c \otimes d) = (ac) \otimes (bd).
\end{gather*}
The operation is clearly symmetric;
\[
(a \otimes b) (c \otimes d) = (ac) \otimes (bd) = (c \otimes d) (a
\otimes b).
\]

\begin{Lemma} \label{lem:D_ProductRule}
The Baker--Hirota operator satisfies a Leibniz rule.  That is
\begin{gather} \label{eq:D_ProductRule}
\mathcal{D}_i\big( (a \otimes b) (c \otimes d) \big) = (a \otimes b)
\mathcal{D}_i(c \otimes d) + (c \otimes d) \mathcal{D}_i(a \otimes
b).
\end{gather}
\end{Lemma}
\begin{proof}
By direct calculation we see that both sides of
(\ref{eq:D_ProductRule}) are
\begin{gather*}
a_ic \otimes bd + ac_i \otimes bd - ac \otimes b_id - ac \otimes
bd_i.  \tag*{\qed}
\end{gather*}
\renewcommand{\qed}{}
\end{proof}

\begin{Corollary} \label{cor:D_LeibnizRule}
The Baker--Hirota operators satisfy a general Leibniz rule,
\begin{gather*} %\label{D_LeibnizRule}
\mathcal{D}_{i_1} \circ \cdots \circ \mathcal{D}_{i_n}\big( (a
\otimes b)(c \otimes d) \big) = \sum_{k=0}^n \sum_{\pi \in \Pi}
\mathcal{D}_{\pi_1} (a \otimes b) \mathcal{D}_{\pi_2}(c \otimes d).
\end{gather*}
Here $\Pi$ is the set of ordered pairs of disjoint partitions,
$\pi$, of the set of indices $I_n$ into subsets, $\pi_1$ and $\pi_2$
of length $n-k$ and $k$ respectively.  The symbol
$\mathcal{D}_{\pi_i}$ represents the concatenation of Baker--Hirota
operators with indices the entries in $\pi_i$.
\end{Corollary}
\begin{proof}
Recall that the standard product rule $(fg)' = f'g + fg'$ leads by
induction to the general Leibniz rule,
\[
(fg)^{(n)}=\sum_{k=0}^n \binom{n}{k} f^{(k)} g^{(n-k)},
\]
where $\binom{n}{k}$ are the binomial coef\/f\/icients.  This leads
identically from Lemma \ref{lem:D_ProductRule} to the following
result for a single Baker--Hirota operator,
\[
\mathcal{D}_i^{n}\big( (a \otimes b)  (c \otimes d) \big) =
\sum_{k=0}^n \binom{n}{k} \mathcal{D}_i^k(a \otimes b)
\mathcal{D}_i^{n-k}(c \otimes d),
\]
where the superscripts on the operator mean it is applied multiple times.  If instead we apply dif\/ferent Baker Hirota operators then we must consider all the dif\/ferent possibilities for $k$ applications of an operator in each step.  This leads to the def\/inition of $\Pi$ as in the corollary.  The number of dif\/ferent possibilities is the number of ways of choosing $k$ from $n$, and so still $\binom{n}{k}$.
\end{proof}

\section[Generalised Baker-Hirota operators]{Generalised Baker--Hirota operators} \label{SEC:H}

In this section we consider a generalisation of the Baker--Hirota
operators which maintains many of the properties from the previous
section.  Such operators will be classif\/ied with an order $m$, and
will be multi-linear, acting on a tensor product of length $m$.  We
f\/irst introduce an operator to dif\/ferentiate within such tensor
products.

\begin{Definition} \label{def:TPdiff}
Def\/ine $\partial_i^{[j]}$ to be an operator which dif\/ferentiates the
$j$th entry in a tensor product with respect to the variable $u_i$,
i.e.
\begin{gather*}
\partial_i^{[j]} \big( f^{[1]} \otimes \dots \otimes f^{[j-1]} \otimes f^{[j]} \otimes f^{[j+1]} \otimes \dots \otimes f^{[m]} \big)\nonumber\\
\qquad{}
= f^{[1]} \otimes \dots \otimes f^{[j-1]} \otimes f_i^{[j]} \otimes
f^{[j+1]} \otimes \dots \otimes f^{[m]}. %\label{eq:TPdiff}
\end{gather*}
\end{Definition}

We note that these operators commute with each other,
\[
\partial_{i_1}^{[j_1]}\partial_{i_2}^{[j_2]} = \partial_{i_2}^{[j_2]}\partial_{i_1}^{[j_1]}.
\]
 For brevity we will often express these tensor products
using a product notation, such as
\[
\bigotimes_{k=1}^m f^{[k]} = f^{[1]} \otimes \dots \otimes f^{[m]}.
\]
We now def\/ine the generalised operators.  We note that these
operators are similar to the (\emph{symmetric operators}) proposed
in~\cite{GRH94}.  However, the authors there considered only the
trilinear cases ($m=3$) in detail, and actually preferred to use a
dif\/ferent non-symmetric basis that could be described as products of
bilinear operators.  Symmetric multilinear Hirota maps of the sort
def\/ined below were discussed in \cite{CA99} in the context of
algebraic covariant and invariant theory.

\begin{Definition} \label{def:H}
Def\/ine $\mathcal{H}_i^{[m]}$ to be the \textit{$m$th order
generalised Baker--Hirota operator}.  This acts on the tensor product
of $m$ functions as
\begin{gather*} %\label{eq:H}
\mathcal{H}_i^{[m]}: \  \bigotimes_{k=1}^m f^{[k]} \longrightarrow
\sum_{j=1}^m   \zeta^{j-1} \partial_i^{[j]} \bigotimes_{k=1}^m
f^{[k]},
\end{gather*}
where $\zeta$ is a primitive $m$th root of unity and $m\geq2$.
\end{Definition}

 We can take $\zeta$ to be any primitive $m$th root of
unity, but we may assume without loss of generality that
\[
\zeta = \exp\left( \frac{2 \pi i}{m}\right).
\]
Recall that
\begin{gather} \label{eq:zeta_sum_zero}
1 + \zeta + \zeta^2 + \dots + \zeta^{m-1} = 0.
\end{gather}
We note that the $2$nd order generalised Baker--Hirota operator is
exactly the original Baker--Hirota operator from Def\/inition
\ref{def:D},
\[
\mathcal{H}_i^{[2]} = \mathcal{D}_i.
\]
For brevity we may write only $\mathcal{H}$-operator unless we mean
to distinguish between the cases.

Since the $\partial_i^{[j]}$-operators commute with each other, it
also clear that they commute with any $\mathcal{H}$-operator,
\[
\partial_{i_1}^{[j]} \circ \mathcal{H}_{i_2}^{[m]} = \mathcal{H}_{i_2}^{[m]} \circ \partial_{i_1}^{[j]}.
\]
From this we can conclude that the $\mathcal{H}$-operators of the
same order commute with each other:
\begin{gather*}
\mathcal{H}_{i_2}^{[m]} \circ \mathcal{H}_{i_1}^{[m]}
 = \mathcal{H}_{i_2}^{[m]} \circ \left(  \sum_{j_1=1}^m   \zeta^{j_1-1} \partial_{i_1}^{[j_1]}
\right) = \sum_{j_1=1}^m   \zeta^{j_1-1} \partial_{i_1}^{[j_1]} \circ \mathcal{H}_{i_2}^{[m]}
= \mathcal{H}_{i_1}^{[m]} \circ \mathcal{H}_{i_2}^{[m]}.
\end{gather*}

We extend the symmetrization operator from Def\/inition \ref{def:S} to
apply it to tensor products of arbitrary length,
\begin{gather*} %\label{eq:Sgen}
S: \  \bigotimes_{k=1}^m f^{[k]} \longrightarrow \prod_{k=1}^m
f^{[k]}.
\end{gather*}
Again, we are usually interested in applying the generalised
operators to a tensor product of the same object.  As before, the
symmetrization operator would annihilate the resulting sum after the
application of a single operator:
\begin{gather*}
S \circ \mathcal{H}_i^{[m]} \left( \bigotimes_{k=1}^m f \right)  =
S\left( \sum_{j=1}^m \, \zeta^{j-1} \partial_i^{[j]}
\bigotimes_{k=1}^m f^{[k]} \right)
= \sum_{j=1}^m \zeta^{j-1} f^{m-1}f_i
= 0,
\end{gather*}
where the f\/inal equality follows from equation
(\ref{eq:zeta_sum_zero}).  In this general case we f\/ind the
symmetrization operator will also annihilate the result after two
applications, so long as $m>2$.  For example, consider applying the
3rd order operator twice:
\begin{gather}
\mathcal{H}_{i_1}^{[3]} \bigotimes_{k=1}^3 f  = \zeta^0 f_{i_1}
\otimes f \otimes f
 + \zeta^1 f \otimes f_{i_1} \otimes f
 + \zeta^2 f \otimes f \otimes f_{i_1},
\nonumber \\
\mathcal{H}_{i_1}^{[3]} \circ \mathcal{H}_{i_2}^{[3]}
\bigotimes_{k=1}^3 f  =   \zeta^0 f_{i_1,i_2} \otimes f \otimes f
 + \zeta^1 f_{i_1} \otimes f_{i_2} \otimes f
 + \zeta^2 f_{i_1} \otimes f \otimes f_{i_2} \nonumber \\
 \hphantom{\mathcal{H}_{i_1}^{[3]} \circ \mathcal{H}_{i_2}^{[3]} \bigotimes_{k=1}^3 f=}{}  + \zeta^1 f_{i_2} \otimes f_{i_1} \otimes f
 + \zeta^2 f \otimes f_{i_1,i_2} \otimes f
 + \zeta^3 f \otimes f_{i_1} \otimes f_{i_2} \nonumber \\
 \hphantom{\mathcal{H}_{i_1}^{[3]} \circ \mathcal{H}_{i_2}^{[3]} \bigotimes_{k=1}^3 f=}{}
  + \zeta^2 f_{i_2} \otimes f \otimes f_{i_1}
 + \zeta^3 f \otimes f_{i_2} \otimes f_{i_1}
 + \zeta^4 f \otimes f \otimes f_{i_1,i_2}.  \label{eq:Hexample}
\end{gather}
So on symmetrizing we have
\[
S \circ \mathcal{H}_{i_1}^{[3]} \circ \mathcal{H}_{i_2}^{[3]}
\bigotimes_{k=1}^3 f = f_{i_1i_2}f^2 \big( \zeta^0 + \zeta^2 +
\zeta^4\big) + f_{i_1}f_{i_2}f\big(\zeta^1+\zeta^2 + \zeta^1+\zeta^3
+ \zeta^2+\zeta^3 \big).
\]
Then since $\zeta^3=1$ we have
\[
S \circ \mathcal{H}_{i_1}^{[3]} \circ \mathcal{H}_{i_2}^{[3]}
\bigotimes_{k=1}^3 f = f_{i_1i_2}f^2 \big( 1 + \zeta^1 +
\zeta^2\big) + 2f_{i_1}f_{i_2}f\big(1+\zeta^1+\zeta^2 \big) = 0.
\]

To see what happens in general we start by considering the ef\/fect of
multiple applications of the operators.

\subsection{Multiple applications of the generalised operators} \label{SEC:H1}

We are not able to use the multiple sum notation of Lemma~\ref{lem:Dntimes}.  Instead we introduce some new notation, which
may seem a little cumbersome at f\/irst, but is useful for proving the
later results.  This notation will be used throughout the rest of
the paper.

Once again let $I_n = \{ i_1, i_2, \dots , i_n \}$ be a set of
indices.  Let $P_n^m$ be a set whose elements $\rho$ are sequences
constructed from the partitions of the number $n$ of length at most
$m$.  If a partition has length less than $m$ then we extend it to
length $m$ by adding zeros.  We denote by $\rho_1,\dots,\rho_m$ the
parts of the partition in decreasing order and then we def\/ine $\rho$
to be the sequence of parts $[\rho_1,\dots,\rho_m]$.  Next we let
$\Psi(\rho)$ be the set of permutations, $\psi$, of the sequence
$\rho$.  We denote by $\psi_1,\dots,\psi_m$ the entries in a permuted
sequence $\psi$.  We def\/ine a function $z(\psi)$ which returns an
integer according to
\begin{gather} \label{eq:z_psi}
z(\psi) =  \sum_{k=1}^m (k-1)(\psi_k).
\end{gather}
Finally we let $\Pi(\psi)$ be the set of all disjoint partitions
$\pi$ of the set of indices $I_n$ into $m$ subsets of lengths given
by the sequence $\psi$.  We denote the resulting subsets of such a
$\pi$ by $\pi_1, \dots, \pi_m$.

\begin{Lemma} \label{lem:Hntimes}
For $n \geq 1$ we have
\begin{gather} \label{eq:Hntimes}
\mathcal{H}_{i_1}^{[m]} \circ \cdots \circ \mathcal{H}_{i_n}^{[m]}
\bigotimes_{k=1}^m f^{[k]} = \sum_{\rho \in P_n^m} \sum_{\psi \in
\Psi(\rho)} \zeta^{z(\psi)} \sum_{\pi \in \Pi(\psi)}
\bigotimes_{k=1}^m f_{\pi_k}^{[k]}.
\end{gather}
\end{Lemma}
\begin{proof}
We do not need to give an inductive proof of Lemma \ref{lem:Hntimes}
as it may be determined directly from a careful analysis of the
resulting dif\/ferential operator acting on the tensor product,
\[
\mathcal{H}_{i_1}^{[m]} \circ \cdots \circ \mathcal{H}_{i_n}^{[m]}
\bigotimes_{k=1}^m f^{[k]} = \left( \prod_{k=1}^{n}
\left(\sum_{j=1}^m \zeta^{j-1} \partial_{i_{k}}^{[j]} \right)
\right)\bigotimes_{k=1}^m f^{[k]}.
\]
The polynomial in $\partial$-operators would be fully symmetric if we had $\zeta=1$.  We organise the terms as described in the lemma so that we collect together those with the same coef\/f\/icient in $\zeta$ and we observe that this coef\/f\/icient is then given using the function in equation (\ref{eq:z_psi}).
\end{proof}

We note that the size of the set $\Pi(\psi)$ is the number of ways
of splitting $n$ elements into $m$ groups of sizes $\psi_1, \dots,
\psi_m$.  This is by def\/inition a multinomial coef\/f\/icient. Further,
since $\psi$ is just a permutation of $\rho$, we have that
\begin{gather*} %\label{eq:size_Pi}
\left| \Pi(\psi) \right| = \left| \Pi(\rho) \right| =
\binom{n}{\rho_1,\dots,\rho_m} = \frac{n!}{\rho_1! \cdots \rho_m!}.
\end{gather*}
We now give an example of the construction to demonstrate the
notation.
Consider the $m=3$, $n=2$ case that we looked at earlier.  Here
$I_n=\{i_1,i_2\}$ and so $P_2^3$ are the partitions of 2 with length
at most 3.  There are two partitions, and we extend them each to
length $3$ to give
\[
\rho^{[1]} = [2,0,0] \qquad \mbox{and} \qquad \rho^{[2]}=[1,1,0].
\]
First we consider $\rho^{[1]}$.  There are three permutations of
$[2,0,0]$ and so three entries in $\Psi(\rho^{[1]})$;
\[
\psi^{[1]} = [2,0,0], \qquad \psi^{[2]} = [0,2,0], \qquad \mbox{and}
\qquad \psi^{[3]}=[0,0,2].
\]
We see that
\[
z(\psi^{[1]}) = (0)(2) + (1)(0) + (2)(0) = 0,
\]
and similarly $z(\psi^{[2]})=2$ and $z(\psi^{[3]})=4$.  For each
$\psi \in \Psi$ there is only one $\pi \in \Pi(\psi)$ since there is
only one possible split of the indices into subsets of these
lengths.  We have
\begin{gather*}
\Pi(\psi^{[1]}) = \{ [\{i_1,i_2\},\{\},\{\}] \}, \qquad
\Pi(\psi^{[2]}) = \{ [\{\},\{i_1,i_2\},\{\}] \}  \\
 \mbox{and}
\qquad \Pi(\psi^{[3]}) = \{ [\{\},\{\},\{i_1,i_2\}] \}.
\end{gather*}
Hence $\rho^{[1]}$ contributes three terms to the sum;
\[
\zeta^0 f^{[1]}_{i_1,i_2} \otimes f^{[2]} \otimes f^{[3]} + \zeta^2
f^{[1]} \otimes f^{[2]}_{i_1,i_2} \otimes f^{[3]} + \zeta^4 f^{[1]}
\otimes f^{[2]} \otimes f^{[3]}_{i_1,i_2}.
\]
Now we consider $\rho^{[2]}$.  There are also three permutations of
$[1,1,0]$ and so again three entries in $\Psi(\rho^{[2]})$;
\[
\psi^{[1]} = [1,1,0], \qquad \psi^{[2]} = [1,0,1], \qquad \mbox{and}
\qquad \psi^{[3]}=[0,1,1].
\]
We calculate $z(\psi^{[1]})=1$, $z(\psi^{[2]})=2$, $z(\psi^{[3]})=3$.
In each of these case there are two possible partitions of the
indices into sets of these lengths.  For the f\/irst we have
\[
\Pi\big(\psi^{[1]}\big) = \{ \, [\{i_1\},\{i_2\},\{\}], \
[\{i_2\},\{i_1\},\{\}] \, \}.
\]
This contributes the terms
\[
\zeta^1 \big( f^{[1]}_{i_1} \otimes f^{[2]}_{i_2} \otimes f^{[3]} +
f^{[1]}_{i_2} \otimes f^{[2]}_{i_1} \otimes f^{[3]} \big)
\]
to the sum.  Similarly we have $ \Pi(\psi^{[2]}) = \{ \,
[\{i_1\},\{\},\{i_2\}], \  [\{i_2\},\{\},\{i_1\}] \, \} $
contributing
\[
\zeta^2 \big( f^{[1]}_{i_1} \otimes f^{[2]} \otimes f^{[3]}_{i_2} +
f^{[1]}_{i_2} \otimes f^{[2]} \otimes f^{[3]}_{i_1} \big)
\]
and $ \Pi(\psi^{[3]}) = \{ \, [\{\},\{i_1\},\{i_2\}], \
[\{\},\{i_2\},\{i_1\}] \, \} $ contributing
\[
\zeta^3 \big( f^{[1]} \otimes f^{[2]}_{i_1} \otimes f^{[3]}_{i_2} +
f^{[1]} \otimes f^{[2]}_{i_2} \otimes f^{[3]}_{i_1} \big).
\]
So we see we have found the nine terms expected from our calculation
in equation~(\ref{eq:Hexample}).

\subsection{Application on the tensor product of a function with itself} \label{SEC:H2}

We will now restrict to the case where the functions in the tensor
product are taken to be the same, and consider the ef\/fect of the
symmetrization operator. In this section we make use of the theory
of symmetric functions.  We gather together the necessary
def\/initions and theory in Appendix \ref{APP_Symm}.

\begin{Lemma} \label{lem:SHntimes}
Let $X=[X_1, \dots ,X_m]$ be given by $X_k = \zeta^{k-1}$.  Then for
$n \geq 1$ we have
\begin{gather} \label{eq:SHntimes}
S \circ \mathcal{H}_{i_1}^{[m]} \circ \cdots \circ
\mathcal{H}_{i_n}^{[m]} \bigotimes_{k=1}^m f = \sum_{\rho \in P_n^m}
M_{\rho}(X)  \sum_{\pi \in \Pi(\rho)} \left(\prod_{k=1}^m
f_{\pi_k}\right),
\end{gather}
where $M$ are the monomial symmetric functions $($see Definition
{\rm \ref{def:mon_sym_fun})}.
\end{Lemma}

\begin{proof}
We consider equation (\ref{eq:Hntimes}) from Lemma
\ref{lem:Hntimes}.  Under the symmetrization operator $S$ the tensor
product becomes a regular product, so
\[
S \left( \sum_{\pi \in \Pi(\psi)} \bigotimes_{k=1}^m f_{\pi_k}^{[k]}
\right) = \sum_{\pi \in \Pi(\psi)}  \prod_{k=1}^m f_{\pi_k}^{[k]}.
\]
Now if all the functions $f^{[k]}=f$ then this sum over $\Pi(\psi)$
is the same for all $\psi \in \Psi(\rho)$.  So we may instead denote
it as a sum over $\Pi(\rho)$ and factor this entire object outside
of the sum over~$\Psi(\rho)$.  We cannot discard the permutations
$\psi$ entirely though, as the term $\zeta^{z(\psi)}$ still varies,
even under symmetrization.  We have
\begin{gather*}
S \circ \mathcal{H}_{i_1}^{[m]} \circ \cdots \circ
\mathcal{H}_{i_n}^{[m]} \bigotimes_{k=1}^m f = \sum_{\rho \in P_n^m}
\left( \sum_{\pi \in \Pi(\rho)} \left(\prod_{k=1}^m f_{\pi_k}\right)
\right) \left( \sum_{\psi \in \Psi(\rho)} \zeta^{z(\psi)} \right).
\end{gather*}
We further analyse the sum in the f\/inal factor.  We have
\[
\sum_{\psi \in \Psi(\rho)} \zeta^{z(\psi)} = \sum_{\psi \in
\Psi(\rho)} \prod_{k=1}^m (\zeta^{k-1})^{\psi_k}.
\]
If we introduce the dummy variables $X_k = \zeta^{k-1}$ then we have
that
\[
\sum_{\psi \in \Psi(\rho)} \zeta^{z(\psi)} = \sum_{\psi \in
\Psi(\rho)} \prod_{k=1}^m X_k^{\psi_k} = M_{\rho}(X_1,\dots,X_m),
\]
from Def\/inition \ref{def:mon_sym_fun}.  Hence the f\/inal factor is identif\/ied as $M_{\rho}(X)$ as described in the lemma.
\end{proof}

 Denote the power sum symmetric functions by $p_k$ (see
Def\/inition \ref{def:power_sums}).  When evaluated on~$X$ as in Lemma
\ref{lem:SHntimes} we f\/ind (see Lemma \ref{lem:zeta_pk}) that
\[
p_k =   \begin{cases} m  & \mbox{if } m|k , \\ 0 &
  \mbox{otherwise.} \end{cases}
\]
By considering expressions in the power sum basis we can determine
when multiple applications of the $\mathcal{H}$-operators map to
zero.
\begin{Corollary} \label{cor:SH_nndivm}
If $m \nmid n$ then
\begin{gather*} %\label{eq:SH_nndivm}
S \circ \mathcal{H}_{i_1}^{[m]} \circ \cdots \circ
\mathcal{H}_{i_n}^{[m]} \bigotimes_{k=1}^m f = 0.
\end{gather*}
\end{Corollary}

\begin{proof}
Every monomial symmetric function is homogeneous in degree.  For
those appearing in Lemma \ref{lem:SHntimes}, the degree is always~$n$.  We may think of the degree as a weight.  Since each power sum
$p_k$ has degree $k$, the expression for $M_{\rho}$ as a polynomial
in $p_1,\dots,p_m$ will only involve terms where the subscripts of
the power sums add to $n$.  Hence if $m \nmid n$ then the expression
cannot contain any terms with $p_m$ as the only variable.  However,
it is only these terms that do not evaluate to zero with the choice
of variables $X_k = \zeta^{k-1}$ by Lemma~\ref{lem:zeta_pk}.

So if $m|n$ then every term in the right hand side of equation (\ref{eq:SHntimes}) is zero.
\end{proof}

\begin{Corollary} \label{cor:SH_ndivm1}
If $m | n$ then $S \circ \mathcal{H}_{i_1}^{[m]} \circ \cdots \circ
\mathcal{H}_{i_n}^{[m]} \bigotimes_{k=1}^m f$ is not identically
zero.
\end{Corollary}

\begin{proof}
Consider the expression in Lemma~\ref{lem:SHntimes}.  One of the terms in equation (\ref{eq:SHntimes}) will correspond the partition on $n$ with $\rho=[n,0,\dots,0]$.  But for such a partition $M_{\rho}=p_{n}$ and by Lemma~\ref{lem:zeta_pk} we know that this power sum is not zero for $m|n$.  So one of the terms in the expression is a~non-zero constant times a monomial in derivatives of $f$ which does not occur in any of the other terms.  Hence the expression cannot be identically zero.
\end{proof}

We can go further and give a formula for the value of the $M_{\rho}$
in the case where $n|m$.  To do this we use a procedure of Doubilet
from~\cite{Doubilet72} to express the monomial symmetric functions
in power sums with coef\/f\/icients given by M\"obius functions,
$\mu(\pi)$.  We give the full details in Appendix~\ref{APP_Symm}
with the key formula stated in equation~(\ref{eq:Mhat_to_pk}).

\begin{Corollary}
The augmented monomial symmetric functions on
$X=[1,\zeta,\dots,\zeta^{m-1}]$ are
\[
\hat{M}_{\rho} = \sum_{\pi \in \Pi'} \mu(\pi) m^{\ell},
\]
where $\ell$ is the number of subsets in $\pi$ and $\Pi'$ contains
only those set partitions $\pi\in\Pi$ such that all parts of
$\bm{\nu(\rho,\pi)}$ are multiples of $m$.
\end{Corollary}
\begin{proof}\sloppy
From Lemma \ref{lem:zeta_pk} we know that when evaluating equation (\ref{eq:Mhat_to_pk}) on the variables \mbox{$X_k = \zeta^{k-1}$}, only those terms such that all parts of $\bm{\nu(\rho,\pi)}$ are multiples of $m$ will be non-zero.  In these cases each $p_i$ is $m$ and so each M\"obius function has a coef\/f\/icient of $m$ to the power of the length of $\bm{\nu(\rho,\pi)}$.
\end{proof}

The monomial symmetric functions involved in Lemma
\ref{lem:SHntimes} will evaluate to the integers for the augmented
functions from the corollary above, divided by the constants from
equation (\ref{eq:Mhat_to_M}).  However, we observe that when $m|n$
then $S \circ \mathcal{H}_{i_1}^{[m]} \circ \cdots \circ
\mathcal{H}_{i_n}^{[m]} \bigotimes_{k=1}^m f$ also has integer
coef\/f\/icients as a polynomial in the derivatives of $f$.  In fact,
these integer coef\/f\/icients all have a common factor of $m$.

\subsection{Leibniz rule} \label{SEC:H3}

We recall the multiplication operation we def\/ined for tensor
products, which multiplies corresponding entries in tensor products
of the same length.  We use it again here to show that the
$\mathcal{H}$-operators all satisfy Leibniz properties.

\begin{Lemma} \label{lem:H_ProductRule}
An $m$th order $\mathcal{H}$-operator satisfies a product rule.
That is
\begin{gather}
\mathcal{H}_i^{[m]}\left( \left(\bigotimes_{k=1}^m f^{[k]}\right)
\left( \bigotimes_{k=1}^m g^{[k]} \right)\right)\nonumber\\
\qquad{} =
\left(\bigotimes_{k=1}^m f^{[k]}\right)
\mathcal{H}_i^{[m]}\left(\bigotimes_{k=1}^m g^{[k]}\right) +
\left(\bigotimes_{k=1}^m g^{[k]}\right)
\mathcal{H}_i^{[m]}\left(\bigotimes_{k=1}^m f^{[k]}\right).\label{eq:H_ProductRule}
\end{gather}
\end{Lemma}

\begin{proof}
Using the def\/initions we have
\begin{gather*}
\mbox{LHS(\ref{eq:H_ProductRule})}  =
\sum_{j=1}^m \, \zeta^{j-1} \partial_i^{[j]} \bigotimes_{k=1}^m
f^{[k]}g^{[k]},
\\
\mbox{RHS(\ref{eq:H_ProductRule})}
= \sum_{j=1}^m \, \zeta^{j-1} \left( \left( \bigotimes_{k=1}^m
f^{[k]} \right)\left( \partial_i^{[j]} \bigotimes_{k=1}^m g^{[k]}
\right) + \left( \bigotimes_{k=1}^m g^{[k]} \right) \left(
\partial_i^{[j]} \bigotimes_{k=1}^m f^{[k]} \right) \right).
\end{gather*}
Comparing the $j$th term in each sum shows them to be equal, (and
hence that the operator $\partial^{[j]}_i$ satisf\/ies its own product
rule).
\end{proof}

\begin{Corollary} \label{cor:H_LeibnizRule}
The $\mathcal{H}$-operators satisfy a general Leibniz rule,
\begin{gather*} %\label{H_LeibnizRule}
\mathcal{H}^{[m]}_{i_1} \circ \cdots \circ \mathcal{H}^{[m]}_{i_n}
\left( \bigotimes_{k=1}^m f^{[k]} \right) \left( \bigotimes_{k=1}^m
g^{[k]} \right) = \sum_{\ell=0}^n \sum_{\pi \in \Pi}
\mathcal{H}^{[m]}_{\pi_1} \left( \bigotimes_{k=1}^m f^{[k]} \right)
\mathcal{H}^{[m]}_{\pi_2}\left( \bigotimes_{k=1}^m g^{[k]} \right).
\end{gather*}
Here $\Pi$ is again the set of disjoint partitions, $\pi$, of the
set of indices $I_n$ into two subsets, $\pi_1$ and $\pi_2$, of
length $n-\ell$ and $\ell$ respectively.  The symbol
$\mathcal{H}^{[m]}_{\pi_i}$ represents the concatenation of
$\mathcal{H}$-operators of the same order with indices the entries
in $\pi_i$.
\end{Corollary}

\begin{proof}
This follows the proof of the original Leibniz rule and the proof of Corollary~\ref{cor:D_LeibnizRule}.
\end{proof}

 We note that as in Corollary \ref{cor:D_LeibnizRule}, the
size of $\Pi$ is still the binomial coef\/f\/icient $\binom{n}{\ell}$.

\section{Abelian functions associated with algebraic curves} \label{SEC:Af}

The f\/irst generalisation of the elliptic functions to hyperelliptic
functions was pioneered by Klein and Baker as described in Baker's
classic texts \cite{ba97,ba07}.  They def\/ined multivariate
$\wp$-functions as discussed in Section \ref{SEC:Intro}.  Hence
these generalised functions are sometimes called Kleinian.  The more
general def\/initions presented on this section are primarily from the
work of Buchstaber, Enolskii and Leykin in \cite{bel97} and the
further developments in \cite{EEL00} and \cite{N10}.

The periodicity conditions of all the higher genus functions are
def\/ined using the period lattice of an underlying algebraic curve.
We use the higher genus $\sigma$-function of the curve to def\/ine
them.  We give explicit def\/initions relating to the class of
$(n,s)$-curves, def\/ined below.  This class is special as the curves
all have a branch point at inf\/inity.  The restriction to this class
is so that we can make use of the well developed theory of series
expansions for the functions associated to such curves to check the
linear independence of functions.

However, we note that one can def\/ine Abelian functions associated to
curves outside this class.  One of the authors has pioneered a so
called \emph{equivariant approach} to the functions in
\cite{CA2008}, where the curve and functions all transform under an
$\mathfrak{sl}_2$ action.  Another recent development is in
\cite{KS12} where the authors def\/ine $\sigma$-functions associated
to Riemann surfaces without making any realisation of the surface as
an algebraic equation.  The def\/initions and results in Sections
\ref{SEC:Qf} and \ref{SEC:Rf} for Abelian functions def\/ined using
$\mathcal{H}$-operators should be applicable in these alternative
theories.  It is only in the calculation of the explicit bases in
Section \ref{SEC:Ex}, that we make use of the $(n,s)$-theory.

\begin{Definition} \label{def:HG_general_curves}
For two coprime integers $(n,s)$ with $s>n$ we def\/ine an
\textit{$(n,s)$-curve}, denoted~$C$, as an algebraic curve
def\/ined by $f(x,y)=0$, where
\begin{gather} \label{eq:general_curve}
f(x,y) = y^n + q_1(x)y^{n-1} + q_{2}(x)y^{n-2} + \cdots +
q_{n-1}(x)y - q_{n}(x).
\end{gather}
Here $x$, $y$ are complex variables and $q_j(x)$ are polynomials in
$x$ of degree (at most) $\lfloor{ js } / n \rfloor$.  We def\/ine a
simple subclass of the curves by setting $q_j(x)=0$ for $0\leq j
\leq n-1$.  These simpler curves are def\/ined using curve parameters $\lambda_j$ as
\begin{gather} \label{eq:ct_curve}
f(x,y) = y^n - (x^s
+\lambda_{s-1}x^{s-1}+\dots+\lambda_{1}x+\lambda_0)
\end{gather}
and are called \textit{cyclic $(n,s)$-curves}.  In the literature the word
`cyclic' is sometimes replaced by `strictly' or `purely $n$-gonal'.
Note that we also have curve parameters for the general $(n,s)$-curves, contained within the $q_j(x)$.
\end{Definition}

Restricting to the cyclic classes results in much easier
computations, but it not usually ne\-cessary for theoretical reasons.
However, the cyclic curves do possess extra symmetry which manifests
itself in the fact that the associated functions satisfy a wider set
of addition formulae.

If the curve $C$ is a non-singular model of $X$ then the genus is
given by $g=\frac{1}{2}(n-1)(s-1)$.  The associated functions will
be multivariate with $g$ variables, $\bm{u} = (u_1, \dots, u_g)$.
For example, a~(2,3)-curve would have genus one and is indeed an
elliptic curve.  The associated Weierstrass~$\sigma$ and
$\wp$-functions depend upon a~single complex variable $u$.

\subsection{Def\/ining the functions} \label{SEC:Af_def}

We f\/irst describe how the period lattice associated to the curve may
be constructed.  We start by choosing a basis for the space of
dif\/ferential forms of the f\/irst kind; the dif\/ferential 1-forms which
are holomorphic on the curve, $C$.  There is a standard procedure to
construct this basis for an $(n,s)$-curve (see for example
\cite{N10}).  In general the basis is given by $h_idx/f_y$, $i=1, \dots,
g$ where the $h_i$ are monomials in $(x,y)$ whose structure may be
predicted by the Weierstrass gap sequence, although other
normalisations are sometimes used.

We next choose a symplectic basis in \(H_1(C,\mathbb{Z})\) of cycles
(closed paths) upon the compact Riemann surface def\/ined by $C$.  We
denote these by $\{\alpha_1, \dots, \alpha_g,$ $\beta_1, \dots,
\beta_g\}$.  We ensure the cycles have intersection numbers
\begin{gather*}
\alpha_i  \alpha_j = 0, \qquad \beta_i \beta_j = 0, \qquad \alpha_i
\beta_j = \delta_{ij} =   \begin{cases}
1 & \mbox{if }  i = j, \\
0 & \mbox{if }  i \neq j.
\end{cases}
\end{gather*}
The choice of these cycles is not unique, but the functions will be
independent of the choice.

We introduce $\bm{dr}$ as a basis of dif\/ferentials of the second
kind.  These are meromorphic dif\/ferentials on $C$ which have their
only pole at $\infty$.  This basis is usually derived alongside the
fundamental dif\/ferential of the second kind.  Rather than repeat the
full details here we refer the reader to~\cite{ba97} for the general
theory and~\cite{bg06} which gives a detailed example construction.

We can now def\/ine the standard period matrices associated to the
curve as
\begin{alignat*}{3}
 &      \omega'  = \left( \oint_{\alpha_k} du_\ell \right)_{k,\ell = 1,\dots,g}, \qquad& &
\qquad\omega'' = \left( \oint_{ \beta_k} du_\ell \right)_{k,\ell = 1,\dots,g}, &  \\
&        \eta'  = \left( \oint_{\alpha_k} dr_{\ell} \right)_{k,\ell = 1,\dots,g}, \qquad & &
\qquad  \eta'' = \left( \oint_{ \beta_k} dr_{\ell} \right)_{k,\ell =
1,\dots,g}&
\end{alignat*}
and def\/ine the period lattice $\Lambda$ formed from $\omega',
\omega''$ by
\[
\Lambda = \big\{ \omega'\bm{m} + \omega''\bm{n}, \  \bm{m},\bm{n}
\in \mathbb{Z}^g \big\}.
\]
This is a lattice in the space $\mathbb{C}^g$.  The Jacobian variety
of $C$ is presented by $\mathbb{C}^g/\Lambda$, and is denoted by
$J$. We def\/ine \(\kappa\) as the modulo \(\Lambda\) map,
\[
\kappa   : \ \mathbb{C}^g \to J.
\]
For $k=1,2, \dots$ def\/ine $\mathfrak{A_k}$, the \emph{Abel map}
from the $k$-th symmetric product \(\mathrm{Sym}^k(C)\) of \(C\)  to
$J$ by
\begin{alignat*}{3}
&\mathfrak{A_k}: \ && \mbox{Sym}^k(C)   \to      J, & \nonumber \\
&&&  (P_1,\dots,P_k)    \mapsto  \left( \int_{\infty}^{P_1} \bm{du} +
\dots + \int_{\infty}^{P_k} \bm{du} \right) \pmod{\Lambda},&
%\label{eq:Abel}
\end{alignat*}
where the $P_i$ are again points upon $C$.  Denote the image of the
$k$-th Abel map by $W^{[k]}$ and def\/ine the \emph{$k$-th standard
theta subset} (sometimes referred to as the $k$-th strata) by
\[
\Theta^{[k]} = W^{[k]} \cup [-1]W^{[k]},
\]
where \([-1]\) means that
\[
[-1](u_1, \dots ,u_g) = (-u_1, \dots ,-u_g).
\]

We are considering functions that are periodic with respect to the
lattice $\Lambda$.
\begin{Definition} \label{def:HG_Abelian}
Let $\mathfrak{M}(\bm{u})$ be a meromorphic function of $\bm{u} \in
\mathbb{C}^g$.  Then $\mathfrak{M}$ is a \textit{standard Abelian
function associated with $C$} if it has poles only along
\(\kappa^{-1}(\Theta^{[g-1]})\) and satisf\/ies, for all $\bm{\ell}\in
\Lambda$,
\begin{gather} \label{eq:HG_Abelian}
\mathfrak{M}(\bm{u} + \bm{\ell}) = \mathfrak{M}(\bm{u}).
\end{gather}
\end{Definition}
All the Abelian functions in this paper are def\/ined using the
$\sigma$-function of the curve.  While not an Abelian function
itself, it does satisfy a quasi-periodicity property.  Let
\(\bm{\delta} = \omega'\bm{\delta'}+\omega''\bm{\delta''}\) be the
Riemann constant with base point $\infty$.  Then $[\bm{\delta}]$ is
the theta characteristic representing the Riemann constant for the
curve~$C$ with respect to the base point $\infty$ and generators
$\{\alpha_j, \beta_j\}$ of $H_1(C,\mathbb{Z})$.  (See for example
\cite[pp.~23--24]{bel97}.)

\begin{Definition} \label{def:HG_sigma}
The \textit{Kleinian $\sigma$-function} associated to a~general
$(n,s)$-curve is def\/ined using a~multivariate $\theta$-function with
characteristic \(\bm{\delta}\) as
\begin{gather*}
\sigma(\bm{u})  =  c \exp \big( \textstyle \frac{1}{2} \bm{u} \eta'
(\omega')^{-1} \bm{u}^T \big)
 \theta[\bm{\delta}]\big((\omega')^{-1}\bm{u}^T \, \big| \, (\omega')^{-1} \omega''\big)
 =  c \exp \big(   \tfrac{1}{2} \bm{u} \eta' (\omega')^{-1}
\bm{u}^T \big) \\
\hphantom{\sigma(\bm{u})  =}{} \times
  \sum_{\bm{m} \in \mathbb{Z}^g} \exp
\bigg[ 2\pi i \big\{
   \tfrac{1}{2} (\bm{m} + \bm{\delta'})^T (\omega')^{-1}
\omega''(\bm{m} + \bm{\delta'}) + (\bm{m} + \bm{\delta'})^T
((\omega')^{-1} \bm{u}^T + \bm{\delta''} )\big\} \bigg].
\end{gather*}
The constant $c$ is dependent upon the curve parameters and the
basis of cycles and is f\/ixed later, following Lemma {\rm
\ref{lem:sigexp}}.
\end{Definition}

We now summarise the key properties of the $\sigma$-function.  See
\cite{bel97} or \cite{N10} for the construction of the
$\sigma$-function to satisfy these properties.  For any point
$\bm{u} \in \mathbb{C}^g$ we denote by $\bm{u}'$ and $\bm{u}''$ the
vectors in $\mathbb{R}^g$ such that
\[
\bm{u}=\omega'\bm{u}'+\omega''\bm{u}''.
\]
Therefore a point  $\bm{\ell}\in\Lambda$ is written as
\[
\bm{\ell} = \omega'\bm{\ell'} + \omega''\bm{\ell''} \in \Lambda,
\qquad \bm{\ell'},\bm{\ell''} \in \mathbb{Z}^g.
\]
For $\bm{u}, \bm{v} \in \mathbb{C}^g$ and $\bm{\ell} \in \Lambda$,
def\/ine $L(\bm{u},\bm{v})$ and $\chi(\bm{\ell})$ as follows:
\begin{gather*}
\hat{L}(\bm{u},\bm{v})  =  \bm{u}^T \big( \eta'\bm{v'} + \eta''\bm{v''} \big), \\
\chi(\bm{\ell})   =  \exp \big[ 2 \pi \mbox{i} \big\{
(\bm{\ell'})^T\delta'' - (\bm{\ell''})^T\delta' +
\tfrac{1}{2}(\bm{\ell'})^T \bm{\ell''} \big\} \big].
\end{gather*}

\begin{Lemma} \label{lem:sig_prop}
Consider the $\sigma$-function associated to an $(n,s)$-curve.
\begin{itemize}\itemsep=0pt
\item It is an entire function on $\mathbb{C}^g$.
\item It has zeros of order one along the set  $\kappa^{-1}(\Theta^{[g-1]})$, commonly called the $\Theta$-divisor.  Further, we have $\sigma(\bm{u}) \neq 0$ outside the set.
\item For all $\bm{u} \in \mathbb{C}^g, \bm{\ell} \in \Lambda$ the function has the quasi-periodicity property:
\begin{gather} \label{eq:sig_per}
\sigma(\bm{u} + \bm{\ell}) = \chi(\bm{\ell})\exp \left[ \hat{L}
\left( \bm{u} + \frac{\bm{\ell}}{2}, \bm{\ell} \right) \right]
\sigma(\bm{u}).
\end{gather}
We let $L=\hat{L}(\bm{u}+\tfrac{\bm{\ell}}{2})$ and $h=\chi e^L$
denote the extra factor introduced.  So $\sigma(\bm{u}+\ell) =
h\sigma(\bm{u})$.
\item It has definite parity given by
\[
\sigma(-\bm{u}) = (-1)^{\frac{1}{24}(n^2-1)(s^2-1)}\sigma(\bm{u}).
\]
\end{itemize}
\end{Lemma}
\begin{proof}
The function is clearly entire from the def\/inition, while the quasi-periodicity and the zeros are classical results (see~\cite{ba97}), that are fundamental to the def\/inition of the function.  They both follow from the properties of the multivariate $\theta$-function.  The parity property is given by Proposition 4(iv) in \cite{N10}.
\end{proof}

We can now def\/ine $\wp$-functions using an analogy of the
relation between $\wp$ and $\sigma$ in the elliptic case.

\begin{Definition} \label{def:nip}
Def\/ine \textit{$m$-index Kleinian $\wp$-functions} for
$m\geq2$ by
\begin{gather*}
\wp_{i_1,i_2,\dots,i_m}(\bm{u}) = - \frac{\partial}{\partial
u_{i_1}} \frac{\partial}{\partial u_{i_2}}\cdots
\frac{\partial}{\partial u_{i_m}} \log \big[ \sigma(\bm{u}) \big],
\end{gather*}
where $i_1 \leq \dots \leq i_m \in \{1,\dots,g\}$.
\end{Definition}
The $m$-index $\wp$-functions are meromorphic with poles of order
$m$ when $\sigma(\bm{u})=0$.  We can check that they satisfy
equation (\ref{eq:HG_Abelian}) and hence they are Abelian by
substituting equation (\ref{eq:sig_per}) directly into their
def\/inition.   The $m$-index $\wp$-functions have def\/inite parity
with respect to the change of variables $\bm{u} \to [-1]\bm{u}$.
This is independent of the underlying curve, with the functions odd
if $m$ is odd and even if $m$ is even.  The ordering of the indices
is irrelevant and so for simplicity we always order in ascending
value.

The new Abelian functions introduce in Sections \ref{SEC:Qf} and
\ref{SEC:Rf} are def\/ined using the operators applied to tensor
products of functions.  We def\/ine an operator for checking their
periodicity.
\begin{Definition} \label{def:LatOp}
Given the lattice $\Lambda$, def\/ine a corresponding \textit{lattice
operator} by
\begin{gather*} %\label{eq:LatOp}
\Lambda_{\ell}: \  \bm{u} \mapsto \bm{u} + \bm{\ell},
\end{gather*}
for $\ell$ some arbitrary point in the lattice.  So
$\Lambda_{\ell}(\sigma) = \sigma(\bm{u} + \ell) = h\sigma$ for
example.  In a slight abuse of notation we also use $\Lambda_{\ell}$
for the operator acting on a tensor product,
\[
\Lambda_{\ell} \bigotimes_{i=1}^n f^{[i]} = \bigotimes_{i=1}^n
\Lambda_{\ell} f^{[i]}.
\]
\end{Definition}

\begin{Lemma} \label{lem:Lam_Com}
The operator $\Lambda_{\ell}$ commutes with $S$, $\mathcal{D}_i$ and
$\mathcal{H}^{[m]}_i$.
\end{Lemma}

 The proofs are all simple by direct calculation.

\subsection{Weights and expansions} \label{SEC:Af_wts}

For a given $(n,s)$-curve we can def\/ine a set of weights, denoted by
$\mathrm{wt}$ and often referred to as the \emph{Sato weights}.  We
start by setting $\mathrm{wt}(x)=-n$, $\mathrm{wt}(y)=-s$ and then
choose the weights of the curve parameters to be such that the curve
equation is homogeneous.  We see that for cyclic curves this imposes
$\mathrm{wt}( \lambda_j )=-n(s-j)$.
The Abel map $\mathfrak{A}_1$ gives an embedding of the curve $C$ upon which we
can def\/ine $\xi= x^{-\frac{1}{n}}$ as the local parameter at the
origin, $\mathfrak{A}_1(\infty)$.  We can then express $y$ and the
basis of dif\/ferentials using $\xi$ and integrate to give series
expansions for $\bm{u}$.  We can check the weights of $\bm{u}$ from
these expansions and see that they are prescribed by the Weierstrass
gap sequence.

By considering Def\/inition \ref{def:nip} we see that the weight of
the $\wp$-functions is the negative of the sum of the weights of the
variables indicated by the indices, irrespective of what the weight
of~$\sigma(\bm{u})$ is.  We note that curves of the same genus will,
notationally, have the same $\wp$-functions, but may exhibit
dif\/ferent behavior as indicated by the dif\/ferent weights of the
variables and functions.  We will discuss the weight of the
$\sigma$-function below.  All other functions discussed are
constructed from $\sigma$ or $\wp$-functions and their weights
follow accordingly.  We can show that all the equations in the
theory are homogeneous in these weights, with a more detailed
discussion of this available, for example, in~\cite{MEe09}.

We can construct a series expansion of the $\sigma$-function about
the origin, as described below.

\begin{Lemma} \label{lem:sigexp}
The Taylor series expansion of $\sigma(\bm{u})$ about the origin may
be written as
\[
\sigma(\bm{u}) = K  SW_{n,s}(\bm{u}) + \sum_{k=0}^{\infty}
C_{k}(\bm{u}).
\]
Here $K$ is a constant, $SW_{n,s}$ the Schur--Weierstrass polynomial
generated by $(n,s)$ and each $C_k$ a~finite, polynomial composed of
products of monomials in $\bm{u}$ of weight $k$ multiplied by
monomials in the curve parameters of weight
$-(\mbox{wt}(\sigma)-k)$.
\end{Lemma}

\begin{proof}
We refer the reader to \cite{N10} for a proof of the relationship between the $\sigma$-function and the Schur--Weierstrass polynomials and note that this was f\/irst discussed in \cite{bel99}.  We see that the remainder of the expansion must depend on the curve parameters and split it up into the dif\/ferent $C_k$ according to the weight split on terms between the parameters and $\bm{u}$.  We can see that each $C_k$ is f\/inite since the number of possible terms with the prescribed weight properties is f\/inite.  In fact, by considering the possible weights of the curve coef\/f\/icients we see that the index $k$ in the sum will, in the cyclic case, actually increase in multiples of $n$.
\end{proof}

The Schur--Weierstrass polynomials are Schur polynomials generated by
a Weierstrass partition, derived in turn from the Weierstrass gap
sequence for $(n,s)$.  See \cite{N10} and \cite{bel99} for more
details on these polynomials.  This connection with the
Schur--Weierstrass polynomials allows us to determine the weight of
the $\sigma$-function as
\[
\mbox{wt}(\sigma) = (1/24)(n^2-1)(s^2-1).
\]
In Def\/inition \ref{def:HG_sigma} we f\/ix $c$ to be the value that
makes the constant $K=1$ in the above lemma.  Some other authors
working in this area may use a dif\/ferent constant and in general
these choices are not equivalent.  However, the constant can be seen
to cancel in the def\/inition of all Abelian functions, leaving
results between the functions independent of $c$.  Note that this
choice of $c$ ensures that the Kleinian $\sigma$-function matches
the Weierstrass $\sigma$-function when the $(n,s)$-curve is chosen
to be the classic elliptic curve.

The expansion can be constructed by considering each $C_k$ in turn,
identifying the possible terms, forming a series with unidentif\/ied
coef\/f\/icients and then determining the coef\/f\/icients by ensuring the
expansions satisfy known properties of the $\sigma$-function.  For
example, using its vanishing properties or the identities between
the associated Abelian functions. Large expansions of this type were
f\/irst introduced in \cite{bg06}, which used the generalised
$\sigma$-function to construct explicit reductions of the Benney
equations.  Since then they have been an integral tool in the
investigation of Abelian functions.  Recently computational
techniques based on the weight properties have been used to derive
much larger expansions and we refer the reader to \cite{MEe09} and~\cite{MEhgt10} for a more detailed discussion of the constructions.

Such expansions are possible for the general $(n,s)$-curves, but the
calculations involved are far simpler for the cyclic cases.  In this
paper such expansions are used only to check linear independence in
the examples of Section~\ref{SEC:Ex}.  The number of $C_k$ required
to do this will depend on the weight of the functions in question.
The expansion will need to go far enough to give the monomials in
curve coef\/f\/icients that may be present in any identity between
functions.

\subsection{Bases of Abelian functions} \label{SEC:Af_bases}

We can classify the Abelian functions according to their pole
structure.  We denote by $\Gamma(m)$ the vector space of Abelian
functions def\/ined upon $J$ which have poles of order at most $m$,
occurring only on $\kappa^{-1}(\Theta^{[g-1]})$, where the $\theta$
and $\sigma$-functions have their zeros and the Abelian functions
their poles.  This is usually called the $\Theta$-divisor.

A key problem is the generation of bases for these vector spaces.
Note that the dimension of the space $\Gamma(m)$ is $m^g$ by the
Riemann--Roch theorem for Abelian varieties (see for example~\cite{la82}).  The f\/irst step in constructing a basis is to include
the entries in the preceding basis for $\Gamma(m-1)$. Subsequently only
functions with poles of order exactly $m$ need to be sought.  The
Kleinian $\wp$-functions are natural candidates and indeed are
suf\/f\/icient to solve the problem in the elliptic case.  Here $g=1$
and so only one new function is required at each stage, which can be
f\/illed by the repeated derivatives of the Weierstrass
$\wp$-functions.

However, if $g>1$ then new classes of functions are required to
complete the bases.  For example, in the genus 2 case we f\/ind that
when considering $\Gamma(3)$ we need an additional entry after
including all possible $\wp$-functions.  The function
\[
\Delta=\wp_{11}\wp_{22} - \wp_{12}^2
\]
is usually taken to f\/ill this hole.  The individual terms in
$\Delta$ have poles of order 4, but when taken together they cancel
to leave poles of order 3 as can easily be checked using Def\/inition~\ref{def:nip}.  We then proceed to consider $\Gamma(4)$ and f\/ind
that two new functions are needed after the inclusion of the 4-index
$\wp$-functions.  The two derivatives of $\Delta$ can play this
role.  Naturally the derivatives of the $\Delta$ have poles of order
four but we must also ensure that they are linearly independent of
the other functions.  This can be checked trivially by noting that
they are of a dif\/ferent weight to both each other and all the other
elements of the basis.

It is simple to check that when considering higher values of $m$,
the preceding basis and its unique derivatives always give the
required number of functions for a new basis, and that they are all
of unique weight and so linearly independent.  Hence the general
basis is as described in Table~\ref{tab:g2Bases}, where $\{ \cdot
\}$ is indicating all functions of this form.

\begin{table}[t]
\caption{Table of bases for Abelian functions associated with a
genus two curve.} \label{tab:g2Bases}
\begin{center}
\begin{tabular}{l | c l}
\hline
\textbf{Space}  &  \textbf{Dimension} & \tsep{1pt}\bsep{1pt} \textbf{Basis} \\ \hline
$\Gamma(0)$      & 1          & \bsep{2pt}   $\{1\}$                                                                               \\
$\Gamma(1)$      & 1          & $\{1\}$                                                                                 \\
$\Gamma(2)$      & 4          & $\{1,\wp_{11},\wp_{12},\wp_{22}\}$                                                     \\
$\Gamma(3)$      & 9          & $\{1,\wp_{11},\wp_{12},\wp_{22}, \wp_{111},\wp_{112}, \wp_{122}, \wp_{222}, \Delta \}$  \\
$\Gamma(4)$      & 16         & $\{1,\dots,\Delta,\wp_{1111},\dots,\wp_{2222}, \partial_1\Delta, \partial_2\Delta \}$   \\
\ $\vdots$         & \ $\vdots$   & \\
$\Gamma(m)$      & $m^2$      & $\{ \mbox{basis for } \Gamma(m-1)\}\cup\{\{\wp_{i_1 \dots i_m}\}, \{\partial_{i_1}\cdots\partial_{i_{m-2}}\Delta\}\}$ \\
\end{tabular}
\end{center}
\end{table}

The genus 1 and 2 cases, in which we get to a stage where new bases
are calculated from the old ones and their derivatives, are special.
They fall into the class where the theta divisor is non-singular and
the $\mathcal{D}$-module structure of such cases is discussed in~\cite{cn08}.
For $g>2$ hyperelliptic
and $g>3$ non-hyperelliptic curves this is not the case and
so new methods are required to derive bases.

One approach to the problem is to def\/ine new functions by matching
poles in algebraic combinations of $\wp$-functions so that they
cancel, analogous to $\Delta$ above.  This method was applied
successfully to derive bases for various vector spaces in
\cite{ME11} and \cite{MEeo11}.  However, this approach does not
generalise easily, which led to the development of the new classes of Abelian functions
in Sections \ref{SEC:Qf} and \ref{SEC:Rf}.

\section[$Q$-functions]{$\boldsymbol{Q}$-functions}  \label{SEC:Qf}

In this section we study the $Q$-functions, def\/ined using the
original Baker--Hirota operator, proving in particular that they are
all Abelian.  The results of this section also act as a motivation
for the more general case discussed in Section~\ref{SEC:Rf}.

\begin{Definition} \label{def:Q}
We def\/ine the \textit{${n}$-index ${Q}$-functions} from the
Kleinian $\sigma$-function as
\begin{gather} \label{eq:Qdef}
Q_{i_1,\dots,i_n} = \left( - \dfrac{1}{2\sigma^2}\right) \left( S
\circ \mathcal{D}_{i_1} \circ \cdots \circ \mathcal{D}_{i_n} \big(
\sigma \otimes \sigma \big)\right), \qquad i_1 \leq \dots  \leq i_n
\in \{1,\dots,g\}.
\end{gather}
\end{Definition}
We note that the indices on the $Q$-functions do not indicate
dif\/ferentiation, but rather the application of operators.  The $n$
in the def\/inition is just the number of operators applied, and is
not related to the $n$ in an $(n,s)$-curve.  Indeed, the results of
this section and the next are independent of the underlying curve.
This def\/inition is equivalent to the following one more commonly
found in the literature, using the Hirota derivative from the
introduction
\[
Q_{i_1, i_2,\dots,i_n}(\bm{u}) =  \frac{(-1)}{2\sigma(\bm{u})^2}
D_{i_1}D_{i_2} \cdots D_{i_n} \sigma(\bm{u}+\bm{v})
\sigma(\bm{u}-\bm{v})   \Big|_{\bm{v}=\bm{0}}.
\]
We prefer Def\/inition \ref{def:Q} as the notation generalises in a
much simpler way.

We use the label $Q$ to follow the notation of Baker in \cite{BA03},
who def\/ined the 4-index functions.  The general def\/inition above was
developed in \cite{MEe09} and \cite{MEhgt10} when 6-index functions
were needed to explicitly solve basis problems.  Baker likely chose
the notation $Q$ since the functions can be viewed as a
generalisation of the 2-index $\wp$-functions.  Comparing with
Def\/inition~\ref{def:nip} we see
\[
Q_{ij} = \frac{\sigma_i\sigma_j - \sigma\sigma_{ij}}{\sigma^2} =
\wp_{ij}.
\]
We aim to prove that the $Q$-functions are Abelian.  So we need to
show that for $\ell \in \Lambda$
\[
Q_{i_1, i_2,\dots,i_n}(\bm{u} + \bm{\ell}) = Q_{i_1,
i_2,\dots,i_n}(\bm{u}).
\]
We can check the periodicity of a 2-index $Q$-function by simply
substituting in the quasi-periodicity property of $\sigma(\bm{u})$
to the expression in $\sigma$-derivatives above.  While this
approach can be used to prove the periodicity of $Q$-functions for
other values of $n$, it does not extend to an inductive proof due to
the symmetrization operator, $S$, in the def\/inition.

Corollary \ref{cor:Dnodd} shows us that the $n$-index $Q$-functions
are zero when $n$ is odd.  A similar calculation shows that for $n$
even the functions are divisible by two.  Lemma \ref{lem:Dntimes}
may be used to give the general formula for an $n$-index
$Q$-function in $\sigma$-derivatives and we may show by induction
that
\begin{gather} \label{eq:SigPerLem}
\sigma_{i_1,i_2,\dots,i_n}(\bm{u} + \bm{\ell}) = \chi e^L
\sum_{m=0}^{n} \left(   \sum_{1 \leq j_1 < j_2 < \dots < j_m
\leq n} \sigma_{I_n
\backslash \{ i_{j_1},i_{j_2},\dots,i_{j_m} \} }
L_{i_{j_1}}L_{i_{j_2}} \cdots L_{i_{j_m}} \right)
\end{gather}
using the fact that second derivatives of $L(\bm{u})$ are zero.
However, substituting these properties into the formula for the
$Q$-functions and observing the cancelations is not trivial.

Instead, we use the properties of the operators to prove the result
using the following property of~$h$, the extra factor introduced in
the quasi-periodicity condition~(\ref{eq:sig_per}).

\begin{Lemma} \label{lem:Dh0}
We have $S \circ \mathcal{D}_{i_1} \circ \cdots \circ
\mathcal{D}_{i_n} (h \otimes h) = 0$ for all $n\geq1$.
\end{Lemma}
\begin{proof}
We note that if $n$ is odd then the result follows directly from
Corollary \ref{cor:Dnodd}.  To show the result also holds for $n$
even consider Lemma \ref{lem:Dntimes} with $f=h$,
\begin{gather} \label{eq:lem:Dhzero}
S \circ \mathcal{D}_{i_1} \!\circ \cdots \circ \mathcal{D}_{i_n} \big(
h \otimes h \big) = \sum_{m=0}^{n}\! ({-}1)^m\!\! \left(
\sum_{1 \leq j_1 < j_2 < \dots < j_m
\leq n}\!\!\! h_{I_n \backslash \{
i_{j_1},i_{j_2},\dots,i_{j_m} \} } h_{i_{j_1},i_{j_2},\dots,
i_{j_m}}\! \right)\!,\!\!\!\!
\end{gather}
Since all the second derivatives of $L$ are zero we have $h_i = \chi
e^{L}L_i$, $h_{ij} = \chi e^{L}L_iL_j$, and by a~simple induction,
\begin{gather}\label{eq:hdiff}
h_{i_1,\dots,i_n} = \chi e^{L} L_{i_1} \cdots L_{i_n}.
\end{gather}
Note that every term in the inner sum in equation
(\ref{eq:lem:Dhzero}) above contains each index in $I_n$ exactly
once.  Hence each of those terms are equal to $\chi^2e^{2L}L_{i_1}
\cdots L_{i_n}$.  The number of terms is the number of ways of
choosing $m$ from~$n$, the binomial coef\/f\/icient.  Hence
\begin{gather*}
S \circ \mathcal{D}_{i_1} \circ \cdots \circ \mathcal{D}_{i_n} \big(
h \otimes h \big)  = \chi^2e^{2L}L_{i_1} \cdots L_{i_n}
\sum_{m=0}^{n} (-1)^m  \binom{n}{m}.
\end{gather*}
The lemma then follows from
\[
\sum_{m=0}^{n} (-1)^m  \binom{n}{m} = 0.  \tag*{\qed}
\]
  \renewcommand{\qed}{}
\end{proof}

We now use this result to give a simple proof of the periodicity of the
$Q$-functions.

\begin{Theorem}
The $n$-index $Q$-functions are all Abelian.
\end{Theorem}
\begin{proof}
By Def\/inition \ref{def:LatOp} of the lattice operator we see that
the Abelian property is equivalent to showing
\[
\Lambda_{\ell} \circ Q_{i_1,\dots,i_n} = Q_{i_1,\dots,i_n}.
\]
Using the def\/inition of the $Q$-functions in equation
(\ref{eq:Qdef}),
Lemma \ref{lem:Lam_Com} which concluded that $\Lambda_{\ell}$
commutes with the other operators, and the periodicity property of
$\sigma$, we have
\begin{gather} \label{eq:Qper_proof1}
\Lambda_{\ell} \circ Q_{i_1,\dots,i_n} = \left( -
\dfrac{1}{2h^2\sigma^2}\right) S \circ \mathcal{D}_{i_1} \circ
\cdots \circ \mathcal{D}_{i_n} \circ \Lambda_{\ell} \big( \sigma
\otimes \sigma \big),
\end{gather}
where $h=\chi e^{L}$.  We observe that under the multiplication
operation for tensor products,
\[
\Lambda_{\ell} (\sigma \otimes \sigma) = (h\sigma \otimes h \sigma)
= (h \otimes h)(\sigma \otimes \sigma).
\]
So we may use the Leibniz rule from Corollary~\ref{cor:D_LeibnizRule} to give
\begin{gather} \label{eq:Qper_proof2}
\mathcal{D}_{i_1} \circ \cdots \circ \mathcal{D}_{i_n}  (h \otimes
h)  (\sigma \otimes \sigma) = \sum_{k=0}^n \sum_{\pi \in \Pi}
\mathcal{D}_{\pi_1} (h \otimes h)  \mathcal{D}_{\pi_2}(\sigma
\otimes \sigma).
\end{gather}
When $k=0$ we see there is only one $\pi \in \Pi$ and that $\pi_1$
is the empty set.  So the inner sum in equation
(\ref{eq:Qper_proof2}) produces just one term,
\[
(h \otimes h) \mathcal{D}_{i_1} \circ \cdots \circ
\mathcal{D}_{i_n}(\sigma \otimes \sigma).
\]
When $k=1$ there are $n$ entries in $\Pi$ and in each case $\pi_1$
is of length one.  Hence the inner sum in equation
(\ref{eq:Qper_proof2}) produces $n$ terms
\[
\mathcal{D}_{i_1} (h \otimes h)  \mathcal{D}_{I_n \backslash
i_1}(\sigma \otimes \sigma) + \mathcal{D}_{i_2} (h \otimes h)
\mathcal{D}_{I_n \backslash i_2}(\sigma \otimes \sigma) + \cdots +
\mathcal{D}_{i_n} (h \otimes h) \mathcal{D}_{I_n \backslash
i_n}(\sigma \otimes \sigma).
\]
Every term here involves a Baker--Hirota operator applied to $(h
\otimes h)$ and hence under symmetrization each term becomes zero by
Lemma \ref{lem:Dh0}.  Similarly, for all $k>0$ the inner sum in
equation (\ref{eq:Qper_proof2}) produces terms which all involve $k$
operators applied to $(h \otimes h)$.  Only the term from $k=0$ will
not vanish under symmetrization.  Hence we have
\begin{gather*} %\label{eq:hsqout}
S \circ \mathcal{D}_{i_1} \circ \cdots \circ \mathcal{D}_{i_n} \big(
(h \otimes h) (\sigma \otimes \sigma)  \big) = h^2 S \circ
\mathcal{D}_{i_1} \circ \cdots \circ \mathcal{D}_{i_n}(\sigma
\otimes \sigma)
\end{gather*}
and so equation (\ref{eq:Qper_proof1}) becomes
\begin{gather*}
\Lambda_{\ell} \circ Q_{i_1,\dots,i_n}
 = \left( - \dfrac{1}{2h^2\sigma^2}\right) h^2 S \circ \mathcal{D}_{i_1} \circ \cdots \circ \mathcal{D}_{i_n}(\sigma \otimes \sigma)
 = Q_{i_1,\dots,i_n}. \tag*{\qed}
\end{gather*}
\renewcommand{\qed}{}
\end{proof}

We have already noted that the 2-index $Q$-function is
exactly the function $\wp_{ij}$.  It is actually possible to express
any $Q$-function as a polynomial in the $\wp$-functions.  In
\cite{eemop07} the authors showed that
\[
Q_{ijk\ell} = \wp_{ijk\ell} - 2
\wp_{ij}\wp_{k\ell}-2\wp_{ik}\wp_{j\ell} -2\wp_{i\ell}\wp_{jk},
\]
and similarly in \cite{MEe09} that
\begin{gather*}
 Q_{ijklmn} = \wp_{ijklmn} - 2\Big[ \big(\wp_{ij}\wp_{klmn} +
\wp_{ik}\wp_{jlmn} + \wp_{il}\wp_{jkmn} + \wp_{im}\wp_{jkln}
 + \wp_{in}\wp_{jklm}\big) \nonumber \\
 \quad{} + \big( \wp_{jk}\wp_{ilmn} + \wp_{jl}\wp_{ikmn} +
\wp_{jm}\wp_{ikln} + \wp_{jn}\wp_{iklm} \big)
+ \big( \wp_{kl}\wp_{ijmn} + \wp_{km}\wp_{ijln} + \wp_{kn}\wp_{ijlm} \big)  \nonumber \\
 \quad {} + \big( \wp_{lm}\wp_{ijkn} + \wp_{ln}\wp_{ijkm} \big) +
\wp_{mn}\wp_{ijkl} \Big]
+ 4\Big[ \big( \wp_{ij}\wp_{kl}\wp_{mn} + \wp_{ij}\wp_{km}\wp_{ln} + \wp_{ij}\wp_{kn}\wp_{lm} \big) \nonumber \\
 \quad {} + \big(\wp_{ik}\wp_{jl}\wp_{mn} + \wp_{ik}\wp_{jm}\wp_{ln} +
\wp_{ik}\wp_{jn}\wp_{lm} \big) + \big(\wp_{il}\wp_{jk}\wp_{mn}  +
\wp_{il}\wp_{jm}\wp_{kn}
+ \wp_{il}\wp_{jn}\wp_{km} \big) \nonumber \\
 \quad {}+ \big(\wp_{im}\wp_{jk}\wp_{ln} + \wp_{im}\wp_{jl}\wp_{kn}  +
\wp_{im}\wp_{jn}\wp_{kl} \big) + \big(\wp_{in}\wp_{jk}\wp_{lm}
+ \wp_{in}\wp_{jl}\wp_{km} + \wp_{in}\wp_{jm}\wp_{kl} \big) \Big].
\end{gather*}
It follows from Theorem \ref{thm:R_to_P} later that the general
formula is
\begin{gather} \label{eq:Q_to_P}
Q_{i_1,\dots,i_n} = \sum_{\rho \in P_n^{[2]}} (-2)^{\ell-1}
\sum_{\pi \in \Pi(\rho)} \wp_{\pi},
\end{gather}
where $P_n^{[2]}$ is the set of partitions of $n$ using only even
numbers and $\ell$ the length of a given partition $\rho$.  As
usual, $\Pi(\rho)$ is the set of all disjoint partitions of the set
of indices into subsets of lengths given $\rho$.  We denote the
resulting subsets of $\pi$ by $\pi_1, \dots, \pi_{\ell}$ and let
$\wp_{\pi}$ be the product of $\wp$-functions with indices the
subsets~$\pi_i$.

\section[$\mathcal{R}$-functions]{$\boldsymbol{\mathcal{R}}$-functions} \label{SEC:Rf}

In this section we study a new class of functions, designed to
generalise the $\wp$ and $Q$-functions in turn.  Hence we have
chosen to call them $\mathcal{R}$-functions.

\begin{Definition} \label{def:R}
We def\/ine the \textit{${n}$-index ${m}$th order
${\mathcal{R}}$-functions} from the Kleinian $\sigma$-function,
using the generalised Baker--Hirota derivatives of Def\/inition~\ref{def:H}, as
\begin{gather} \label{eq:Rdef}
\mathcal{R}^{[m]}_{i_1,\dots,i_n} = \left( -
\dfrac{1}{m\sigma^m}\right) \left( S \circ \mathcal{H}^{[m]}_{i_1}
\circ \cdots \circ \mathcal{H}^{[m]}_{i_n} \bigotimes_{k=1}^m \sigma
\right),
\end{gather}
where $i_1 \leq \dots  \leq i_n \in \{1,\dots,g\}.$
\end{Definition}
Like the $Q$-functions, the indices of $\mathcal{R}$-functions do
not indicate dif\/ferentiation, but rather the application of
operators.

Corollary \ref{cor:SH_nndivm} shows us that the $n$-index
$\mathcal{R}$-functions are zero unless $n$ is a multiple of $m$.
We can also observe from Corollary \ref{cor:SH_ndivm1} that for such
$n$ the functions are not identically zero.  From the def\/inition it
is clear that the $m$th order $\mathcal{R}$-functions have poles of
order at most $m$, occurring on the $\Theta$-divisor.  The main
result of this section will be a proof that these functions are also
Abelian, and hence can be used to complete the basis for
$\Gamma(m)$.  Following the approach taken for the $Q$-functions in
the last section, we will do this using the Leibniz rule for the
operators and the ef\/fect of the operators applied to $h$, the extra
factor introduced in the quasi-periodicity condition
(\ref{eq:sig_per}) for the $\sigma$-function.

\begin{Lemma} \label{lem:Hh0}
We have, for all $n\geq1$,
\[
S \circ \mathcal{H}^{[m]}_{i_1} \circ \cdots \circ
\mathcal{H}^{[m]}_{i_n} \bigotimes_{k=1}^m h = 0.
\]
\end{Lemma}
\begin{proof}
We note that if $n$ is not a multiple of $m$ then the result follows
directly from Corollary~\ref{cor:SH_nndivm}.
In Lemma~\ref{lem:Dh0} we could show directly that the
symmetrization of the original Baker--Hirota operators evaluated on
$h$ gave zero due to well known properties of the binomial
coef\/f\/icient involved.  We saw in Section \ref{SEC:H2} that for the
general operators the coef\/f\/icients involve not just multinomial
coef\/f\/icients but also M\"obius functions and so we use a dif\/ferent
approach.

First we recall Lemma \ref{lem:Hntimes}, which allows us to conclude
that
\begin{gather} \label{lem:Hh0_proof1}
\mathcal{H}^{[m]}_{i_1} \circ \cdots \circ \mathcal{H}^{[m]}_{i_n}
\bigotimes_{k=1}^m h = \sum_{\ell_n} \left( c_{\ell_n}
\bigotimes_{k=1}^m h_{\pi_{k,\ell}} \right).
\end{gather}
Here $\ell_n$ is an index and the sum is f\/inite with the number of
terms depending on $n$.  Each~$\pi_{k,\ell}$ is a~subset of the
indices $I_n=\{i_1,\dots,i_n\}$ such that
$\pi_{1,\ell},\dots,\pi_{k,\ell}$ are a disjoint partition of~$I_n$.
We further recall equation~(\ref{eq:hdiff}) which stated that for
the function $h$ we have $h_{i_1,\dots,i_n} = \chi e^{L} L_{i_1}
\cdots L_{i_n}$.  Hence each of these tensor products will symmetrize
to give the same function,
\[
S \left( \bigotimes_{k=1}^m  h_{\pi_{k,\ell}} \right) =
\prod_{k=1}^m h_{\pi_{k,\ell}} = \chi^m e^{mL}L_{i_1} \cdots L_{i_n}.
\]
So we have
\begin{gather} \label{lem:Hh0_proof2}
S \circ \mathcal{H}^{[m]}_{i_1} \circ \cdots \circ
\mathcal{H}^{[m]}_{i_n} \bigotimes_{k=1}^m h = \chi^m e^{mL}L_{i_1}
\cdots L_{i_n} \left( \sum_{\ell_n} c_{\ell_n} \right).
\end{gather}

We aim to show that the sum of constants $c_{\ell_n}$ in equation
(\ref{lem:Hh0_proof2}) is always zero, which would allow us to
conclude the lemma.  For the case $n=1$ we have
\[
\mathcal{H}^{[m]}_{i_1} \bigotimes_{k=1}^m h = \sum_{j=1}^m
\zeta^{j-1} \partial^{[j]}_{i_1} \bigotimes_{k=1}^m h,
\]
and so $\ell_1$ runs from 1 to $m$ with
$c_{\ell_1}=\zeta^{\ell_1-1}$.  Clearly the sum of these constants
is zero since $\zeta$ is an $m$th root of unity.

Now consider applying another Hirota operator to equation
(\ref{lem:Hh0_proof1}):
\begin{gather*}
\mathcal{H}^{[m]}_{i_1} \circ \cdots \circ
\mathcal{H}^{[m]}_{i_{n+1}} \bigotimes_{k=1}^m h
= \sum_{j=1}^m \zeta^{j-1} \partial_{i_{n+1}}^{[j]} \sum_{\ell_n}
c_{\ell_n} \bigotimes_{k=1}^m h_{\pi_k} = \sum_{j=1}^m \zeta^{j-1}
\sum_{\ell_n} c_{\ell_n} \partial_{i_{n+1}}^{[j]} \bigotimes_{k=1}^m
h_{\pi_k}.
\end{gather*}
Since
\[
h_{\pi_k,i_{n+1}} = L_{i_{n+1}}h_{\pi_k}
\]
we have that for all $j=1,\dots,m$,
\[
S\left( \partial_{i_{n+1}}^{[j]} \bigotimes_{k=1}^m h_{\pi_k}
\right) = \chi^m e^{mL}L_{i_1} \cdots L_{i_{n+1}}.
\]
So we see that
\[
S \circ \mathcal{H}^{[m]}_{i_1} \circ \cdots \circ
\mathcal{H}^{[m]}_{i_{n+1}} \bigotimes_{k=1}^m h = \chi^m
e^{mL}L_{i_1} \cdots L_{i_{n+1}} \left( \sum_{j=1}^m \zeta^{j-1}
\sum_{\ell_n} c_{\ell_n} \right).
\]
Hence
\[
\sum_{\ell_{n+1}} c_{\ell_{n+1}} = \sum_{j=1}^m \zeta^{j-1}
\sum_{\ell_n} c_{\ell_n} = 0. \tag*{\qed}
\]
\renewcommand{\qed}{}
\end{proof}

We can now prove the periodicity of the $\mathcal{R}$-functions.

\begin{Theorem}
The $n$-index $m$th order $\mathcal{R}$-functions are all Abelian.
\end{Theorem}
\begin{proof}
By Def\/inition \ref{def:LatOp} of the lattice operator we see that
the Abelian property is equivalent to showing
\[
\Lambda_{\ell} \circ \mathcal{R}^{[m]}_{i_1,\dots,i_n} =
\mathcal{R}^{[m]}_{i_1,\dots,i_n}.
\]
Using the def\/inition of the $\mathcal{R}$-functions in equation
(\ref{eq:Rdef}), Lemma \ref{lem:Lam_Com} which concluded that
$\Lambda_{\ell}$ commutes with the other operators, and the
periodicity property of $\sigma$, we have
\begin{gather} \label{eq:Rper_proof1}
\Lambda_{\ell} \circ \mathcal{R}^{[m]}_{i_1,\dots,i_n} = \left( -
\dfrac{1}{mh^m\sigma^m}\right) S \circ \mathcal{H}^{[m]}_{i_1} \circ
\cdots \circ \mathcal{H}^{[m]}_{i_n} \circ \Lambda_{\ell}
\bigotimes_{k=1}^m \sigma,
\end{gather}
where $h=\chi e^{L}$.  We observe that under the multiplication
operation for tensor products
\[
\Lambda_{\ell} \left( \bigotimes_{k=1}^m \sigma \right) =
\bigotimes_{k=1}^m h\sigma = \left( \bigotimes_{k=1}^m h \right)
\left( \bigotimes_{k=1}^m \sigma \right).
\]
So we may use the Leibniz rule from Corollary
\ref{cor:H_LeibnizRule} to give
\begin{gather} \label{eq:Rper_proof2}
\mathcal{H}^{[m]}_{i_1} \circ \cdots \circ \mathcal{H}^{[m]}_{i_n}
\bigotimes_{k=1}^m h  \bigotimes_{k=1}^m \sigma = \sum_{\ell=0}^n
\sum_{\pi \in \Pi} \mathcal{H}^{[m]}_{\pi_1} \left(
\bigotimes_{k=1}^m h \right) \mathcal{H}^{[m]}_{\pi_2} \left(
\bigotimes_{k=1}^m \sigma\right).
\end{gather}
When $\ell=0$ we see there is only one $\pi \in \Pi$ and that
$\pi_1$ is the empty set.  So the inner sum in equation
(\ref{eq:Rper_proof2}) produces just one term,
\[
\left( \bigotimes_{k=1}^m h \right) \mathcal{H}^{[m]}_{i_1} \circ
\cdots \circ \mathcal{H}^{[m]}_{i_n} \left(\bigotimes_{k=1}^m
\sigma\right).
\]
For all $\ell>0$ the inner sum in equation (\ref{eq:Qper_proof2})
produces terms which all involve $\ell$ operators applied to
$\otimes_{k=1}^m h$.  So from Lemma \ref{lem:Hh0}, only the term
from $\ell=0$ will not vanish under symmetrization.  Hence we have
\begin{gather*} %\label{eq:hsqout2}
S \circ \mathcal{H}^{[m]}_{i_1} \circ \cdots \circ
\mathcal{H}^{[m]}_{i_n} \left( \bigotimes_{k=1}^m h \right) \left(
\bigotimes_{k=1}^m \sigma\right) = h^m S \circ
\mathcal{H}^{[m]}_{i_1} \circ \cdots \circ
\mathcal{H}^{[m]}_{i_n}\left(\bigotimes_{k=1}^m \sigma\right)
\end{gather*}
and so equation (\ref{eq:Rper_proof1}) becomes
\begin{gather*}
\Lambda_{\ell} \circ \mathcal{R}^{[m]}_{i_1,\dots,i_n}
 = \left( - \dfrac{1}{mh^m\sigma^m}\right) h^m S \circ \mathcal{H}_{i_1} \circ \cdots \circ \mathcal{H}_{i_n}\bigotimes_{k=1}^m \sigma
 = R^{[m]}_{i_1,\dots,i_n}.  \tag*{\qed}
\end{gather*}
\renewcommand{\qed}{}
\end{proof}

Since the $\mathcal{R}$-functions are Abelian we know they can be
written algebraically using the Kleinian $\wp$-function.  We make
this relationship explicit in the next theorem.
\begin{Theorem} \label{thm:R_to_P}
The $n$-index $m$th order $\mathcal{R}$-functions may be expressed
in Kleinian $\wp$-functions as
\begin{gather} \label{eq:R_to_P}
\mathcal{R}^{[m]}_{i_1,\dots,i_n} = \sum_{\rho \in P_n^{[m]}}
(-m)^{\ell-1} \sum_{\pi \in \Pi(\rho)} \wp_{\pi},
\end{gather}
where $P_n^{[m]}$ is the set of partitions of $n$ using only
integers divisible by $m$ and $\ell$ the length of a~given partition
$\rho$.  As usual, $\Pi(\rho)$ is the set of all disjoint partitions
of the set of indices into subsets of lengths given $\rho$.  We
denote the resulting subsets of $\pi$ by $\pi_1, \dots, \pi_{\ell}$
and let $\wp_{\pi}$ be the product of $\wp$-functions with indices
the subsets $\pi_i$.
\end{Theorem}

 For example, in the $m=3$ case we have
\begin{gather*}
\mathcal{R}^{[3]}_{i_1,i_2,i_3,i_4,i_5,i_6}  =
\wp_{i_1,i_2,i_3,i_4,i_5,i_6}
- 3\big( \wp_{i_1,i_2,i_3,}\wp_{i_4,i_5,i_6} + \wp_{i_1,i_2,i_4}\wp_{i_3,i_5,i_6} + \wp_{i_1,i_2,i_5}\wp_{i_3,i_4,i_6} \\
\hphantom{\mathcal{R}^{[3]}_{i_1,i_2,i_3,i_4,i_5,i_6}  =}{}
 + \wp_{i_1,i_2,i_6}\wp_{i_3,i_4,i_5}
+ \wp_{i_1,i_3,i_4}\wp_{i_2,i_5,i_6} + \wp_{i_1,i_3,i_5}\wp_{i_2,i_4,i_6} + \wp_{i_1,i_3,i_6}\wp_{i_2,i_4,i_5} \\
\hphantom{\mathcal{R}^{[3]}_{i_1,i_2,i_3,i_4,i_5,i_6}  =}{}
 + \wp_{i_1,i_4,i_5}\wp_{i_2,i_3,i_6} +
\wp_{i_1,i_4,i_6}\wp_{i_2,i_3,i_5} +
\wp_{i_1,i_5,i_6}\wp_{i_2,i_3,i_4} \big).
\end{gather*}
Note that equation (\ref{eq:R_to_P}) clearly reduces to equation
(\ref{eq:Q_to_P}) in the $m=2$ case.

\begin{proof}
In this proof we def\/ine a symbol $\hat{\wp}$ such that
$\sigma=e^{\hat{\wp}}$.  Then $\hat{\wp}=\log(\sigma)$,
$\hat{\wp}_{i}=\sigma_i/\sigma$ and
\[
\hat{\wp}_{i_1,\dots,i_n} = - \wp_{i_1,\dots,i_n}
\]
for $n\geq2$.  It is important to note that $\hat{\wp}$ is not the
Weierstrass $\wp$-function and neither $\hat{\wp}$ nor
$\hat{\wp_{i}}$ are Abelian functions. We consider an $m$th order
$\mathcal{H}$-operator acting on a tensor product of $\sigma$.
\[
\mathcal{H}_{i_1}^{[m]} \bigotimes_{k=1}^m \sigma =
\mathcal{H}_{i_1}^{[m]} \bigotimes_{k=1}^m e^{\hat{\wp}} =
\sum_{j=1}^m   \zeta^{j-1} \partial_i^{[j]} \bigotimes_{k=1}^m
e^{\hat{\wp}}.
\]
This will be a sum of terms, each a tensor product in which every
entry has a factor $e^{\hat{\wp}}$.  So using the multiplication for
tensor products, we have
\begin{gather} \label{eq:R2Pproof1}
\mathcal{H}_{i_1}^{[m]} \bigotimes_{k=1}^m \sigma = \left(
\sum_{j=1}^m   \zeta^{j-1} \left( 1 \otimes 1 \otimes \dots \otimes
\hat{\wp}_{i_1} \otimes 1 \otimes \dots \otimes 1\right)\right)
\left( \bigotimes_{k=1}^m e^{\hat{\wp}} \right)
\end{gather}
with the $\hat{\wp}_{i_1}$ occurring in the $j$th position.

For any subset $I \in I_n=\{i_1,\dots,i_n\}$ def\/ine
\[
\Sigma_{I} = \sum_{j=1}^m   \zeta^{|I|(j-1)} \left( 1 \otimes 1
\otimes \dots \otimes \hat{\wp}_{I} \otimes 1 \otimes \dots \otimes
1\right).
\]
Then equation (\ref{eq:R2Pproof1}) becomes
\begin{gather} \label{eq:R2Pproof2}
\mathcal{H}_{i_1}^{[m]} \left( \bigotimes_{k=1}^m \sigma \right) =
\Sigma_{\{i_1\}} \left( \bigotimes_{k=1}^m \sigma \right).
\end{gather}
Now consider applying a second operator.  Using the product rule for
$\mathcal{H}$-operators from Lem\-ma~\ref{lem:H_ProductRule} we have
\begin{gather} \label{eq:R2Pproof3}
\mathcal{H}_{i_1}^{[m]} \circ \mathcal{H}_{i_2}^{[m]} \left(
\bigotimes_{k=1}^m \sigma \right)  =
\mathcal{H}_{i_2}^{[m]}(\Sigma_{\{i_1\}})\left( \bigotimes_{k=1}^m
\sigma \right) + \Sigma_{\{i_1\}} \mathcal{H}_{i_2}^{[m]}\left(
\bigotimes_{k=1}^m \sigma \right).
\end{gather}
Applying the $\mathcal{H}$-operator on $\Sigma_{\{i_1\}}$ will produce a
sum of terms, many of which will have zero as one of the entries in
the tensor product.  In this proof we will be applying operators and
then symmetrizing, meaning that these terms will all vanish in the
end and so we need not keep track of them. We let the symbol
$\doteq$ indicate not a true equality, but that the dif\/ference
consists only of terms with zero as one of the entries in the tensor
product.  When we apply the $\mathcal{H}$-operator on $\Sigma_{I}$
the only terms which will not vanish are those where the
dif\/ferentiation is applying to the entry with $\hat{\wp}_{I}$.  We
hence have
\[
\mathcal{H}_{i_k}^{[m]} ( \Sigma_{I} ) \doteq \Sigma_{\{I,i_k\}}.
\]
This result, along with (\ref{eq:R2Pproof2}), simplify equation
(\ref{eq:R2Pproof3}) to
\begin{gather*}
\mathcal{H}_{i_1}^{[m]}\circ\mathcal{H}_{i_2}^{[m]} \left(
\bigotimes_{k=1}^m \sigma \right)  \doteq \Sigma_{\{i_1,i_2\}} \left(
\bigotimes_{k=1}^m \sigma \right) + \Sigma_{ \{i_1\} }\Sigma_{ \{i_2\} } \left(
\bigotimes_{k=1}^m \sigma \right).
\end{gather*}
An induction gives
\begin{gather} \label{eq:R2Pproof4}
\mathcal{H}_{i_1}^{[m]}\circ \dots \circ \mathcal{H}_{i_n}^{[m]}
\left( \bigotimes_{k=1}^m \sigma \right) \doteq \sum_{\rho \in P_n}
\sum_{\pi \in \Pi(\rho)} \Sigma_{\pi} \left( \bigotimes_{k=1}^m
\sigma \right),
\end{gather}
where $P_n^{[m]}$ is the set of partitions of $n$ and the other
notation as stated in the theorem.

Now consider the application of the symmetrization operator.  We
have
\[
S\left( \bigotimes_{k=1}^m \sigma \right) = \sigma^m \qquad
\mbox{and} \qquad S(\Sigma_{I}) = \hat{\wp}_{I} \left(\sum_{j=1}^m
\, \zeta^{|I|(j-1)})\right).
\]
As in the proof of Lemma \ref{lem:zeta_pk}, if $m$ divides $|I|$
then we obtain
\[
S(\Sigma_{I}) = m\hat{\wp}_I = -m\wp_{I}.
\]
and obtain zero otherwise.  Hence (\ref{eq:R2Pproof4}) becomes
\[
S \circ \mathcal{H}_{i_1}^{[m]}\circ \dots \circ
\mathcal{H}_{i_n}^{[m]} \left( \bigotimes_{k=1}^m \sigma \right) =
\sum_{\rho \in P_n^{[m]}} (-m)^{\ell} \sum_{\pi \in \Pi(\rho)}
\wp_{\pi} \sigma^m
\]
and dividing by $-m\sigma^m$ gives the result of the theorem.
\end{proof}

It is clear from their def\/initions that the $Q$-functions are the
2nd order $\mathcal{R}$-functions:
\[
Q_{i_1, i_2,\dots,i_n}(\bm{u}) =
\mathcal{R}^{[2]}_{i_1,i_2,\dots,i_n}(\bm{u}).
\]
Hence in turn the two-index $\wp$-functions are the 2-index 2nd
order $\mathcal{R}$-functions:
\[
\wp_{i,j}(\bm{u})=\mathcal{R}^{[2]}_{i,j}(\bm{u}).
\]
However, we see from the following corollary that in fact all the
Kleinian $\wp$-functions are themselves $\mathcal{R}$-functions.

\begin{Corollary} \label{cor:nPisR}
The $m$-index $m$th order $\mathcal{R}$-functions are exactly the
$m$-index $\wp$-functions,
\begin{gather*}%\label{eq:nPisR}
\wp_{i_1,i_2,\dots,i_m}(\bm{u}) =
\mathcal{R}^{[m]}_{i_1,i_2,\dots,i_m}(\bm{u}).
\end{gather*}
\end{Corollary}

\begin{proof}
This is the application of Theorem \ref{thm:R_to_P} in the case
$n=m$.
\end{proof}

Recall that the $n$-index $\wp$-functions had def\/inite
parity indicated by $n$; i.e.\ they are odd if $n$ is odd and even if
$n$ is even.  This property extends to the $\mathcal{R}$-functions,
and hence the $Q$-functions also.

\begin{Lemma} \label{lem:Rparity}
The $n$-index $\mathcal{R}$-functions have parity equal to that of
$n$, with respect to the change of variables $\bm{u} \to
[-1]\bm{u}$.
\end{Lemma}
The proof is simple but lengthy by considering the dif\/ferent
combinations of parity of the components.  The full details are in
Appendix \ref{App_Rparity}.

\section{New bases of Abelian functions} \label{SEC:Ex}

In this section we present some new bases of Abelian functions,
completed using $\mathcal{R}$-functions.  These are for the two
canonical genus three cases.  In this section we work with functions of only
three variables and so we drop the commas between indices for brevity.  Since the genus is f\/ixed so are the
dimensions of corresponding bases.  However the actual bases
themselves require dif\/ferent functions, made apparent by the
dif\/ferent weights in the two cases (see Section \ref{SEC:Af_wts}).
We note that like the $\wp$-functions, the weights of the
$\mathcal{R}$-functions are the sum of the weights of the variables
indicated by the indices.  This can be seen by considering
$\mathcal{R}$-functions as a sum of terms, each with $\sigma^m$ in
the denominator and a monomial of $\sigma$-derivatives in the
numerator, which includes all the indices.  The weight of
$\sigma$-derivatives is the weight of $\sigma$ minus the weight of
the indices and in each term the weight of $\sigma$ cancels leaving
the overall weight determined only by the indices.

For each $\Gamma(m)$ a basis is constructed systematically as
follows.  We start by including the basis for the preceding case so
we then need only search for functions with poles of order $m$.  We
consider a set of possible basis functions, say
$\mathcal{R}$-functions with a given number of indices, and proceed
in ascending weight order.  We use the $\sigma$-expansion for the
corresponding $(n,s)$-curve to see if a given function is linearly
independent of the existing basis entries and if so add it to the
basis.  It is important to include the functions from the preceding
bases when testing linear independence, since although the $m$th
order $\mathcal{R}$-functions have poles of order at most $m$, they
may in fact have lower order poles when associated with certain
curves and this can only be checked using the expansion.  Once the
set has been exhausted, we consider another set if necessary.

We use the weight theory to consider only linear combinations of
functions (and curve parameters) which give the correct weight.  We
then substitute in suf\/f\/icient terms of the expansion and see if
there are solutions for the coef\/f\/icients.  Such calculations can be
lengthly and grow in CPU time and memory requirement with both the
genus and the number of poles.  Signif\/icant computational
simplif\/ications can be made by writing procedures that take further
advantage of the weight structure present in the theory.  For
example, when expanding the product of series it is only necessary
to multiply those terms which will give the correct f\/inal weight, as
the other terms must all cancel.  We can further reduce calculations
by noting that there is a f\/inite weight range that basis entries can
take.  Entries in $\Gamma(m)$ can have weight no lower than
$-m\mbox{wt}(\sigma)$ (see Lemma 3.4 in \cite{MEeo11}).

The bases presented below are all for functions associated with
general $(n,s)$-curves as in equation (\ref{eq:general_curve}).
However, we only need use the series expansions associated to the
corresponding cyclic curves in equation (\ref{eq:ct_curve}) to check
linear independence.  If an element cannot be expressed using the
basis with the restriction on the parameters, then neither will it
be expressible with the wider set of parameters.  Further, we only
need to use suf\/f\/icient expansion to give non-zero evaluations of the
functions considered in order to check whether they are linearly
independent.

We note that while these bases are not unique, (they can formulated
using dif\/ferent functions), their weight structures are.  Also,
there are other invariants of the bases, such as the number of odd
and even functions.  We have several conjectures on these
invariants, which are still under investigation.

\subsection{Hyperelliptic curves}

The hyperelliptic curve of genus 2 is a special case as its bases
can be f\/initely generated by dif\/ferentiation (see Table
\ref{tab:g2Bases}).  If desired, they could be reformulated using
$\mathcal{R}$-functions.  All the $\wp$-functions are themselves
$\mathcal{R}$-functions and the extra function, $\Delta \in
\Gamma(3)$, may be replaced by the six index $\mathcal{R}$-function,
$\mathcal{R}^{[3]}_{122222}$.  Of course, $\Delta$ may be expressed
as a linear combination of the basis entries.  In the cyclic case
the relationship is,
\[
\Delta = -\frac{2}{5}\lambda_{4}\mathcal{R}^{[2]}_{11} +
\frac{4}{5}\lambda_{4}^2\mathcal{R}^{[2]}_{12} -
\frac{7}{5}\lambda_{3} \mathcal{R}^{[2]}_{12} -
\frac{1}{20}\mathcal{R}^{[3]}_{122222} - \frac{9}{5}\lambda_{1}.
\]
In the genus 3 hyperelliptic case we f\/ind that a basis for
$\Gamma(2)$ is
\begin{gather} \label{eq:g3H_Gamma2}
\big\{1, \mathcal{R}^{[2]}_{11}, \mathcal{R}^{[2]}_{12},
\mathcal{R}^{[2]}_{13}, \mathcal{R}^{[2]}_{22},
\mathcal{R}^{[2]}_{23}, \mathcal{R}^{[2]}_{33},
\mathcal{R}^{[2]}_{2222} \big\}.
\end{gather}
We note that $\mathcal{R}^{[2]}_{2222}$ is a second order
$\mathcal{R}$-function, and so equal to $Q_{2222}$ as def\/ined in
Section \ref{SEC:Qf}.  We note further that there are two other
4-index $Q$-functions of this weight; $Q_{1133}$ and $Q_{1223}$ and
either of these may be used instead.  Indeed, there are even more alternative
$Q$-functions that could play the role if we allow more
indices.

We now proceed to give a basis for $\Gamma(3)$.  We arrange the
elements in the set as vertical lists purely to make the structure
(number of functions of each type) clearer to the reader
\begin{gather} \label{eq:g3H_Gamma3}
\left\{ \begin{array}{ccccccc} (\ref{eq:g3H_Gamma2}),   &
\mathcal{R}^{[3]}_{113}, & \mathcal{R}^{[3]}_{133}, &
\mathcal{R}^{[3]}_{233}, &
\mathcal{R}^{[3]}_{233333}, & \mathcal{R}^{[3]}_{222222}, & %\mathcal{R}^{[3]}_{133333333}, \\
\partial_{1}\mathcal{R}^{[2]}_{2222}, \\
\mathcal{R}^{[3]}_{111}, & \mathcal{R}^{[3]}_{122}, &
\mathcal{R}^{[3]}_{222}, & \mathcal{R}^{[3]}_{333}, &
\mathcal{R}^{[3]}_{133333}, & \mathcal{R}^{[3]}_{113333}, & %\mathcal{R}^{[3]}_{123333333}, \\
\partial_{2}\mathcal{R}^{[2]}_{2222}, \\
\mathcal{R}^{[3]}_{112}, & \mathcal{R}^{[3]}_{123}, &
\mathcal{R}^{[3]}_{223}, &                          &
\mathcal{R}^{[3]}_{222333}, & \mathcal{R}^{[3]}_{112333}, & %\mathcal{R}^{[3]}_{113333333}
\partial_{3}\mathcal{R}^{[2]}_{2222}
\end{array} \right\}.
\end{gather}
Here $(\ref{eq:g3H_Gamma2})$ is representing the eight entries in
the basis for $\Gamma(2)$ presented in equation
(\ref{eq:g3H_Gamma2}). We note that the three f\/irst derivatives of
$\mathcal{R}^{[2]}_{2222}$ could be replaced by 9-index third order
$\mathcal{R}$-functions if desired. Finally, a basis for $\Gamma(4)$
is
\begin{gather*}
\left\{ \begin{array}{ccccccccc}
(\ref{eq:g3H_Gamma3}),    & \mathcal{R}^{[4]}_{1223}, & \partial_{3}\mathcal{R}^{[3]}_{233333}, & \partial_{2}\mathcal{R}^{[3]}_{113333}, & \mathcal{R}^{[4]}_{12333333}, \\
\mathcal{R}^{[4]}_{1111}, & \mathcal{R}^{[4]}_{1233}, & \partial_{2}\mathcal{R}^{[3]}_{233333}, & \partial_{1}\mathcal{R}^{[3]}_{113333}, & \mathcal{R}^{[4]}_{11333333}, \\
\mathcal{R}^{[4]}_{1112}, & \mathcal{R}^{[4]}_{1333}, & \partial_{3}\mathcal{R}^{[3]}_{133333}, & \partial_{2}\mathcal{R}^{[3]}_{112333}, & \mathcal{R}^{[4]}_{11233333}, \\
\mathcal{R}^{[4]}_{1113}, & \mathcal{R}^{[4]}_{2222}, & \partial_{1}\mathcal{R}^{[3]}_{233333}, & \partial_{3}\mathcal{R}^{[3]}_{222222}, & \mathcal{R}^{[4]}_{12223333}, \\
\mathcal{R}^{[4]}_{1122}, & \mathcal{R}^{[4]}_{2223}, & \partial_{2}\mathcal{R}^{[3]}_{133333}, & \partial_{1}\mathcal{R}^{[3]}_{112333}, & \mathcal{R}^{[4]}_{11133333}, \\
\mathcal{R}^{[4]}_{1123}, & \mathcal{R}^{[4]}_{2233}, & \partial_{1}\mathcal{R}^{[3]}_{133333}, & \partial_{2}\mathcal{R}^{[3]}_{222222}, & \mathcal{R}^{[4]}_{22222222}, \\
\mathcal{R}^{[4]}_{1133}, & \mathcal{R}^{[4]}_{2333}, & \partial_{2}\mathcal{R}^{[3]}_{222333}, & \partial_{1}\mathcal{R}^{[3]}_{222222}, & \mathcal{R}^{[4]}_{11123333}  \\
\mathcal{R}^{[4]}_{1222}, & \mathcal{R}^{[4]}_{3333}, &
\partial_{1}\mathcal{R}^{[3]}_{222333}, &
&
\end{array} \right\}.
\end{gather*}
In this case those f\/irst derivatives of third order
$\mathcal{R}$-functions could not be replaced by any fourth order
$\mathcal{R}$-functions.  We may conclude this because previous
calculations in \cite{MEeo11} have shown that the basis must contain
some odd functions with poles of order 4.  However, since the number
of indices on the $4$th order $\mathcal{R}$-functions must be a
multiple of 4, all these functions must be even by Lemma
\ref{lem:Rparity}.  The odd functions with poles of order 4 are all
found within the derivatives of the previous basis.

Although we have not proved that all bases in this case can be
constructed this way, we have overcome one of the key dif\/f\/iculties
in constructing such bases, which becomes apparent when examining
the weights.  In this case the variables $\bm{u}=(u_1,u_2,u_3)$ have
weights (5,3,1) while the $\sigma$-function has weight $6$.  A basis
for $\Gamma(m)$ must have an entry with the minimal weight of
$-m\mbox{wt}(\sigma)$ (see Lemma 3.4 in \cite{MEeo11}).  So in this
case each $\Gamma(m)$ requires a function of weight $-6m$ in its
basis and such a function cannot be a derivative of a function in
$\Gamma(m-1)$.  This is since such a function could have weight no
lower than
\[
-6(m-1)-5 = -6m+1
\]
achieved by dif\/ferentiating a minimal weight function in
$\Gamma(m-1)$ by $u_1$.  We can now always choose the minimal weight
function in the basis of $\Gamma(m)$ to be the $2m$-index function
$\mathcal{R}^{[m]}_{22\dots2}$.

\subsection{Trigonal curves}

The trigonal curves are those $(n,s)$-curves with $n=3$.  The
simplest is the $(3,4)$-curve which also has genus three, but is
non-hyperelliptic.  We f\/ind that a basis for $\Gamma(2)$ is
\begin{gather} \label{eq:g3T_Gamma2}
\big\{1, \mathcal{R}^{[2]}_{11}, \mathcal{R}^{[2]}_{12},
\mathcal{R}^{[2]}_{13}, \mathcal{R}^{[2]}_{22},
\mathcal{R}^{[2]}_{23}, \mathcal{R}^{[2]}_{33},
\mathcal{R}^{[2]}_{2222} \big\}.
\end{gather}
We note that symbolically, this looks the same as equation
(\ref{eq:g3H_Gamma2}).  However, the two theories dif\/fer in the
weights of the functions.  In the hyperelliptic case
$\bm{u}=(u_1,u_2,u_3)$ had weight (5,3,1) while in the trigonal case
the weights are $(5,2,1)$.

Once again, $\mathcal{R}^{[2]}_{2222}=Q_{2222}$ and there were other
choices of $Q$-function.  This time there was only one alternative
4-index $Q$-function of this weight, $Q_{1333}$, contrary to the
alternatives in the hyperelliptic case, making the symbolic
equivalence somewhat coincidental.  However, there is also a
structural equivalence in the use of six 2-index functions and one
4-index function.

For $\Gamma(3)$ in the trigonal case we f\/ind the basis
\begin{gather} \label{eq:g3T_Gamma3}
\left\{ \begin{array}{ccccccc} (\ref{eq:g3T_Gamma2}),   &
\mathcal{R}^{[3]}_{113}, & \mathcal{R}^{[3]}_{133}, &
\mathcal{R}^{[3]}_{233}, &
\mathcal{R}^{[3]}_{333333}, & \mathcal{R}^{[3]}_{113333}, & %\mathcal{R}^{[3]}_{133333333}, \\
\partial_{1}\mathcal{R}^{[2]}_{2222}, \\
\mathcal{R}^{[3]}_{111}, & \mathcal{R}^{[3]}_{122}, &
\mathcal{R}^{[3]}_{222}, & \mathcal{R}^{[3]}_{333}, &
\mathcal{R}^{[3]}_{222333}, & \mathcal{R}^{[3]}_{122333}, & %\mathcal{R}^{[3]}_{233333333}, \\
\partial_{2}\mathcal{R}^{[2]}_{2222}, \\
\mathcal{R}^{[3]}_{112}, & \mathcal{R}^{[3]}_{123}, &
\mathcal{R}^{[3]}_{223}, &                          &
\mathcal{R}^{[3]}_{133333}, & \mathcal{R}^{[3]}_{122233}, & %\mathcal{R}^{[3]}_{333333333}
\partial_{3}\mathcal{R}^{[2]}_{2222}, \\
\end{array} \right\}.
\end{gather}
So the symbolic equivalence is now lost; for example the function
$\mathcal{R}^{[3]}_{333333}$ is present in this basis but can not
replace any function in the basis (\ref{eq:g3H_Gamma3}).  However,
the structural equivalence is still present; i.e. in both the
hyperelliptic and trigonal cases the number of functions with a
given number of indices was the same.  Similarly, for $\Gamma(4)$ in
the trigonal case we f\/ind the basis
\begin{gather*} %\label{eq:g3T_Gamma4}
\left\{ \begin{array}{ccccccccc}
(\ref{eq:g3T_Gamma3}),    & \mathcal{R}^{[4]}_{1223}, & \partial_{3}\mathcal{R}^{[3]}_{333333}, & \partial_{2}\mathcal{R}^{[3]}_{122333}, & \mathcal{R}^{[4]}_{22333333}, \\
\mathcal{R}^{[4]}_{1111}, & \mathcal{R}^{[4]}_{1233}, & \partial_{2}\mathcal{R}^{[3]}_{333333}, & \partial_{1}\mathcal{R}^{[3]}_{133333}, & \mathcal{R}^{[4]}_{22233333},  \\
\mathcal{R}^{[4]}_{1112}, & \mathcal{R}^{[4]}_{1333}, & \partial_{3}\mathcal{R}^{[3]}_{222333}, & \partial_{2}\mathcal{R}^{[3]}_{122233}, & \mathcal{R}^{[4]}_{13333333}, \\
\mathcal{R}^{[4]}_{1113}, & \mathcal{R}^{[4]}_{2222}, & \partial_{1}\mathcal{R}^{[3]}_{333333}, & \partial_{2}\mathcal{R}^{[3]}_{113333}, & \mathcal{R}^{[4]}_{12233333}, \\
\mathcal{R}^{[4]}_{1122}, & \mathcal{R}^{[4]}_{2223}, & \partial_{2}\mathcal{R}^{[3]}_{222333}, & \partial_{1}\mathcal{R}^{[3]}_{122333}, & \mathcal{R}^{[4]}_{12223333}, \\
\mathcal{R}^{[4]}_{1123}, & \mathcal{R}^{[4]}_{2233}, & \partial_{2}\mathcal{R}^{[3]}_{133333}, & \partial_{1}\mathcal{R}^{[3]}_{122233}, & \mathcal{R}^{[4]}_{11333333}, \\
\mathcal{R}^{[4]}_{1133}, & \mathcal{R}^{[4]}_{2333}, & \partial_{3}\mathcal{R}^{[3]}_{122333}, & \partial_{1}\mathcal{R}^{[3]}_{113333}, & \mathcal{R}^{[4]}_{11223333}  \\
\mathcal{R}^{[4]}_{1222}, & \mathcal{R}^{[4]}_{3333}, &
\partial_{1}\mathcal{R}^{[3]}_{222333}, &
&
\end{array} \right\}.
\end{gather*}
Again, the number of functions of each type is the same.  We note
that although bases for these spaces have been derived before in
\cite{eemop07} and \cite{MEeo11}, this structural similarity only
becomes apparent now, when using $\mathcal{R}$-functions.

We conjecture that in general a basis for $\Gamma(m)$ can always be
constructed using the preceding basis, $m$th order
$\mathcal{R}$-functions and the f\/irst derivatives of $(m-1)$th order
$\mathcal{R}$-functions, which would solve the problem of
identifying enough suitable basis functions.

\appendix

\section{Proof of Lemma \ref{lem:Dntimes}} \label{APP_lemDntimes}

We need to verify equation (\ref{eq:Dntimes}),
\begin{gather*}
\mathcal{D}_{i_1} \circ \cdots \circ \mathcal{D}_{i_n} \big( f
\otimes g \big) = \sum_{m=0}^{n} (-1)^m \left( \,
\sum_{1 \leq j_1 < j_2 < \dots < j_m \leq n} f_{I_n \backslash \{
i_{j_1},i_{j_2},\dots,i_{j_m} \} } \otimes g_{i_{j_1},i_{j_2},\dots,
i_{j_m}} \right).
\end{gather*}
\begin{proof}
We give an inductive proof.  First we observe the case $n=1$ where
\begin{gather*}
\mbox{RHS}(\ref{eq:Dntimes})  = (-1)^0\left( f_{I_1 \backslash
\varnothing} \otimes g_{\varnothing} \right)
+   (-1)^1 \left( \sum_{j_1=1}^1 f_{I_1 \backslash \{i_{j_1}\}} \otimes g_{i_{j_1}} \right) \\
\hphantom{\mbox{RHS}(\ref{eq:Dntimes})}{}  = f_{I_1} \otimes g_{\varnothing} - f_{\varnothing} \otimes g_{i_1} =
f_{i_1}\otimes g - f \otimes g_{i_1}
\end{gather*}
as required.
We now assume the result holds for $n=k$ and consider the case
$n=k+1$.
\begin{gather}
 \mathcal{D}_{i_1} \circ \cdots \circ \mathcal{D}_{i_{k+1}} \big( f
\otimes g \big)
= \mathcal{D}_{i_{k+1}} \circ \mathcal{D}_{i_1} \cdots \circ \mathcal{D}_{i_k} \big( f \otimes g \big)  \nonumber \\
  \qquad {} = \mathcal{D}_{i_{k+1}} \left( \sum_{m=0}^{k} (-1)^m
\left(   \sum_{1 \leq j_1 < j_2 < \dots < j_m \leq k}  f_{I_k \backslash \{
i_{j_1},i_{j_2},\dots,i_{j_m} \} } \otimes g_{i_{j_1},i_{j_2},\dots
,i_{j_m}} \right) \right)
\nonumber \\
 \qquad{} = \sum_{m=0}^{k} (-1)^m \left( \left(
\sum_{1 \leq j_1 < j_2 < \dots < j_m \leq k}
f_{I_{k+1} \backslash \{ i_{j_1},i_{j_2},\dots,i_{j_m} \} }\otimes g_{i_{j_1},i_{j_2},\dots,i_{j_m}} \right) \right. \nonumber \\
\left.  \qquad \quad{} -
\left(   \sum_{1 \leq j_1 < j_2 < \dots < j_m \leq k} f_{I_{k} \backslash \{
i_{j_1},i_{j_2},\dots,i_{j_m} \} }\otimes
g_{i_{j_1},i_{j_2},\dots,i_{j_m},i_{k+1}} \right)\right).
\label{eq:Dntimesproof}
\end{gather}
We need to show that equation (\ref{eq:Dntimesproof}) leads to
\begin{gather} \label{eq:Dntimesproof2}
\mathcal{D}_{i_1}\! \circ\! \cdots\! \circ \mathcal{D}_{i_{k+1}}\! \big( f
\!\otimes\! g \big) \! =\! \sum_{m=0}^{k+1} \!(-1)^m \!\!\left(
\sum_{1 {\leq} j_1 {<} j_2 {<}\cdots {<} j_m {\leq} k{+}1} \!\!\!\!\! f_{I_{k+1} \backslash \{
i_{j_1},i_{j_2},\dots,i_{j_m} \} } \!\!\otimes\!
g_{i_{j_1},i_{j_2},\dots,i_{j_m}}\! \right)\!.\!\!\!\!\!\!
\end{gather}
Each value of $m$ in the outer sum in equation
(\ref{eq:Dntimesproof}) gives two parts.  The f\/irst part will
contribute to one value of $m$ in equation (\ref{eq:Dntimesproof2})
with the second contributing to the following value.  These two
parts are required to f\/ill each value of $m$ in equation~(\ref{eq:Dntimesproof2}), with the exception of the f\/irst and last
terms in the sum, for which one part is suf\/f\/icient.

Consider the contribution of the f\/irst part of equation
(\ref{eq:Dntimesproof}) when $m=0$.  We obtain $f_{I_{k+1}}\otimes
g$ which is what we require for $m=0$ in equation
(\ref{eq:Dntimesproof2}).

Similarly consider the contribution of the second part of equation
(\ref{eq:Dntimesproof}) when $m=k$.  We obtain $(-1)^{k+1}f \otimes
g_{I_{k+1}}$ which is what we require for $m=k+1$ in equation
(\ref{eq:Dntimesproof2}).

Now we consider the general term.  We combine the contribution of
the f\/irst part of equation~(\ref{eq:Dntimesproof}) when $m=m$ with
the contribution of the second part when $m=m-1$ to obtain
\begin{gather*}
 (-1)^m \left(   \sum_{1 \leq j_1 < j_2 < \dots < j_m \leq k} f_{I_{k+1} \backslash
\{ i_{j_1},i_{j_2},\dots,i_{j_m} \} } \otimes
g_{i_{j_1},i_{j_2},\dots, i_{j_m}}
\right. \\
\left.\qquad{} + \sum_{1 \leq j_1 < j_2 < \dots < j_m \leq k} f_{I_{k}
\backslash \{ i_{j_1},i_{j_2},\dots,i_{j_{m-1}} \} } \otimes
g_{i_{j_1},i_{j_2},\dots, i_{j_{m-1}},i_{k+1}} \right).
\end{gather*}
Note that the minus sign in the second part was combined with the
$(-1)^{m-1}$ outside to factor a $(-1)^m$ out altogether.

We can check that the terms here are referring to the same partition
of the set of indices over the tensor product.  Counting in the
f\/irst part we see we have $k+1-m$ indices and $m$ indices.
Similarly in the second part we have $k-(m-1)$ indices and $m-1+1$
indices, the same values.

Now note that the terms in the second part never contain $i_{k+1}$
on the left of the tensor product, but always contain it on the
right.  Hence these terms are those extra ones we would obtain if we
ran the sum in the f\/irst part to $k+1$ instead of $k$.  So these
terms can be combined into the one sum to give,
\[
(-1)^m \left(   \sum_{1 \leq j_1 < j_2 < \dots < j_m \leq k+1} f_{I_{k+1}
\backslash \{ i_{j_1},i_{j_2},\dots,i_{j_m} \} } \otimes
g_{i_{j_1},i_{j_2},\dots, i_{j_m}} \right),
\]
the term in equation (\ref{eq:Dntimesproof2}) when $m=m$.
\end{proof}

\section{Symmetric functions} \label{APP_Symm}

We gather together some def\/initions and results for symmetric
functions which we require.

\begin{Definition} \label{def:mon_sym_fun}
The \textit{monomial symmetric polynomials} in $m$ variables,
$x=[x_1,\dots,x_m]$ are each def\/ined by a sequence of natural
numbers of length $m$, say $\rho=[\rho_1,\dots,\rho_m]$, as
\begin{gather*} %\label{eq:mon_sym_fun}
M_{\rho}(x) = \sum_{\psi \in \Psi(\rho)} x^{\psi}.
\end{gather*}
Following the notation introduced for Lemma \ref{lem:Hntimes},
$\Psi(\rho)$ is the set of all permutations of $\rho$ with entries
denoted by $\rho=[\rho_1,\dots,\rho_m]$.  We then let $x^{\psi}$
denote a corresponding monomial,
\[
x^{\psi} = x_1^{\psi_1}x_2^{\psi_2} \cdots x_m^{\psi_m}.
\]
We def\/ine the \textit{augmented monomial symmetric polynomials} as
\begin{gather*} %\label{eq:aug_mon_sym_fun}
\hat{M}_{\rho}(x) = \sum_{\psi \in \Psi(x)} \psi^{\rho}.
\end{gather*}
This time the sum is over permutations of the variables rather than
permutations of the powers.
\end{Definition}

 For example,
\begin{gather*}
M_{[3,1,1]}(x_1,x_2,x_3)  = x_1^3x_2x_3 + x_1x_2^3x_3 + x_1x_2x_3^3, \\
\hat{M}_{[3,1,1]}(x_1,x_2,x_3)  = x_1^3x_2x_3 + x_1^3x_3x_2 + x_2^3x_1x_3 + x_2^3x_3x_1 + x_3^3x_1x_2 + x_3^3x_2x_1 \\
\hphantom{\hat{M}_{[3,1,1]}(x_1,x_2,x_3)}{} = 2M_{[3,1,1]}(x_1,x_2,x_3).
\end{gather*}
Let $\eta_i$ be the \emph{multiplicity} of $i \in \rho$, i.e.\ the
number of entries in $\rho$ equal to $i$.  Then we have
\begin{gather} \label{eq:Mhat_to_M}
\hat{M}_{\rho}(x) = \left(\prod_{i} \eta_i!\right) M_{\rho}(x).
\end{gather}

The monomial symmetric functions naturally occur when considering
the application of $\mathcal{H}$-operators in Section \ref{SEC:H2}.
They occur evaluated on the roots of unity.  Symmetric polynomials
in the roots of unity have been considered previously in~\cite{Lascoux85}, with nice results for some of the other classes.
For example, the elementary symmetric polynomials reduce to
\[
e_1 = 0, \ e_2=0 , \ \dots, \ e_{m-1}=0, \ e_m = (-1)^{m+1}.
\]
However, in our case, the power sum basis is the most
useful.

\begin{Definition} \label{def:power_sums}
The \textit{Newton power sums} in $m$ variables $x_1, \dots, x_m$
are def\/ined for $k\geq0$ as
\[
p_k(x_1,\dots,x_m)=x_1^k + x_2^k + \cdots +x_m^k.
\]
\end{Definition}
Any symmetric polynomial in $m$ variables can be expressed as a
polynomial with rational coef\/f\/icients in $p_1,\dots,p_m$. We observe
how they evaluate on our choice of variables.

\begin{Lemma} \label{lem:zeta_pk}
The power sums evaluated on $X=[X_1, \dots ,X_m]$ where $X_k =
\zeta^{k-1}$ for $\zeta$ a~pri\-mi\-tive $m$th root of unity satisfy
\[
p_k =  \begin{cases} m  & \mbox{if } m|k, \\ 0 &
  \mbox{otherwise.} \end{cases}
\]
\end{Lemma}

\begin{proof}
We have
\begin{gather*}
p_k(X)=  = 1 + \zeta ^{k} + \zeta^{2k} + \cdots + \zeta^{(m-1)k} =
\frac{1-\zeta^{km}}{1-\zeta^k}.
\end{gather*}
Since $\zeta$ is an $m$th root of unity, the numerator is always
zero.  The denominator is only non-zero for $m|k$.  Hence if $k$ is
not a multiple of $m$ then $p_k$=0. If $k=m$ then we have
\[
p_k = X_1^m + X_2^m + \cdots + X_m^m = 1^m + \zeta^{m} + \zeta^{2m} +
\cdots +\zeta^{(m-1)m} = m.
\]
Similarly if $m|k$ then we have a sum of $m$ terms, each $\zeta$ raised to a power divisible by $m$.  Hence the polynomial evaluates to $m$.
\end{proof}

We now introduce a procedure for expressing symmetric functions in
power sums.  Although every symmetric function of $m$ variables may
be expressed as a polynomial in the f\/irst $m$ power sums $p_k$, this
explicit procedure will instead give an expression in the f\/irst~$n$
power sums, where~$n$ is the number partitioned by the sequence
$\rho=[\rho_1,\dots,\rho_m]$.

We again let $\Pi$ be a set of partitions of indices.  This time it
will be all disjoint parti\-tions,~$\pi$, of the set of indices
$I_m=\{i_1,\dots,i_m\}$ into subsets denoted by $\pi_1, \dots,
\pi_{\ell}$.  Note that we do not restrict to a given number of
subsets or specify their size.  We see that $\ell$ will range from~1, (when~$\pi$ has just $\pi_1=I_m$ itself), to~$m$ (when~$\pi$
consists of the $m$ subsets with one index each).  We def\/ine the
associated M\"obius function $\mu(\pi)$ by
\begin{gather*} %\label{eq:Mobius}
\mu(\pi) = (-1)^{m-\ell} \prod_{i=1}^{\ell} (|\pi_{i}| - 1)!
\end{gather*}
and refer readers to \cite{Rota64} for more information on such
functions.

Next we def\/ine length $\ell$ vectors
$\bm{\nu}=[\nu_1,\dots,\nu_{\ell}] \in \mathbb{N}^{\ell}$ by
\begin{gather*} %\label{eq:rhopi}
\bm{\nu}(\bm{\rho},\bm{\pi}) = \left[ \sum_{i_j \in \pi_1} \rho_j, \dots,
\sum_{i_j \in \pi_\ell} \rho_j \right].
\end{gather*}
Finally we def\/ine the function $p_{\bm{\nu}}$ as the product of
power sum functions whose indices are determined by the values in
the vector $\bm{\nu}$,
\[
p_{\bm{\nu}} = p_{\nu_1} \cdots p_{\nu_2}.
\]
We can now state the following formula of Doubilet from
\cite{Doubilet72},
\begin{gather} \label{eq:Mhat_to_pk}
\hat{M}_{\rho} = \sum_{\pi \in \Pi} \mu(\pi) p_{\bm{\nu(\rho,\pi)}}.
\end{gather}
So the augmented monomial symmetric functions have an integer
polynomial expression in the power sums $p_1,\dots,p_n$. The
expression for the monomial symmetric functions will be the same but
with a fractional multiplicative constant depending on the
multiplicities of $\rho$ as described in equation
(\ref{eq:Mhat_to_M}).

We give an example in the case $m=3$.  We consider the disjoint
partitions of $I_3 = \{i_1,i_2,i_3\}$.  The set $\Pi$ has f\/ive
elements:
\begin{alignat*}{5}
& \pi^{[1]}: \quad && \pi^{[1]}_1= \{i_1,i_2,i_3\}, \qquad && && &                       \\
& \pi^{[2]}: \quad && \pi^{[2]}_1= \{i_1,i_2\},   \qquad && \pi^{[2]}_2=\{i_3\},  &&& \\
& \pi^{[3]}: \quad && \pi^{[3]}_1= \{i_1,i_3\},  \qquad  && \pi^{[3]}_2=\{i_2\},  &&& \\
& \pi^{[4]}: \quad && \pi^{[4]}_1= \{i_2,i_3\},  \qquad  && \pi^{[4]}_2=\{i_1\},  &&& \\
& \pi^{[5]}: \quad && \pi^{[5]}_1= \{i_1\}        \qquad &&
  \pi^{[5]}_2=\{i_2\}, \qquad&& \pi^{[5]}_3=\{i_3\}. &
\end{alignat*}
The M\"obius function then gives
\begin{gather*}
\mu(\pi^{[1]})  = (-1)^{3-1}  2!=+2,     \\
\mu(\pi^{[2]}) = \mu(\pi^{[3]}) = \mu(\pi^{[4]})  = (-1)^{3-2}   1!0!=-1,  \\
\mu(\pi^{[5]})  = (-1)^{3-3}   (0!)^3=+1.
\end{gather*}
Hence any augmented monomial symmetric function in $3$ variables may
be expressed using the power sums in 3 variables as
\[
\hat{M}_{\rho} = 2p_{\rho_1+\rho_2+\rho_3} -
p_{\rho_1+\rho_2}p_{\rho_3} - p_{\rho_1+\rho_3}p_{\rho_2} -
p_{\rho_2+\rho_3}p_{\rho_1} - p_{\rho_1}p_{\rho_2}p_{\rho_3}.
\]
For example,
\begin{gather*}
\hat{M}_{[3,1,1]} = 2p_5 - 2p_4p_1 - p_2p_3 + p_3p_1^2.
\end{gather*}
Note that when we have entries in $\rho$ which are zero then we need
to recall that $p_0=m$.  So in the $m=3$ case we have, for example,
\begin{gather*}
\hat{M}_{[3,0,0]} = 2p_3 - 3p_3p_0 + p_3p_0^2 = 2p_3.
\end{gather*}

\section{Proof of Lemma \ref{lem:Rparity}} \label{App_Rparity}

We prove that the $\mathcal{R}$-functions have parity equal to that
of $n$, the number of indices.

\begin{proof}
Think of the $n$-index $m$th order $\mathcal{R}$-functions as
\begin{gather} \label{lem:Rparity_proof}
\mathcal{R}^{[m]}_{i_1,\dots,i_n} = \frac{\mbox{polynomial in
$\sigma$-derivatives}}{\sigma^m}.
\end{gather}
Each term in the polynomial in the numerator is a monomial in
$\sigma$ and its derivatives containing $n$-indices overall (see
Lemmas \ref{lem:Hntimes} and \ref{lem:SHntimes}).  Recall that the
$\sigma$-function is either odd or even.  In the case of
$(n,s)$-curves this can be determined from $(n,s)$ (see Lemma~\ref{lem:sig_prop}).  We refer to derivatives of $\sigma$ with
respect to an odd number of variables as $\sigma_{\rm o}$-derivatives and
similarly $\sigma_{\rm e}$-derivatives if dif\/ferentiated with respect to
an even number of variables.

Suppose f\/irst that $\sigma$ is an even function and so the
denominator of (\ref{lem:Rparity_proof}) is even.  In this case the
$\sigma_{\rm o}$-derivatives are odd and the $\sigma_{\rm e}$-derivatives even.
If $n$ is odd then each term in the numerator of
(\ref{lem:Rparity_proof}) must contain an odd number of
$\sigma_{\rm o}$-derivatives.  Similarly if $n$ is even then each term
would have an even number.  Hence the numerator, and therefore
$\mathcal{R}^{[m]}_{i_1,\dots,i_n}$, has parity the same as $n$.

Now suppose that $\sigma$ is an odd function.  This time the
denominator of (\ref{lem:Rparity_proof}) has parity equal to that of
$m$, and it is the $\sigma_{\rm e}$-derivatives which are odd while the
$\sigma_{\rm o}$-derivatives are even.  In this case we need to be more
careful.  Note that each term in the numerator of
(\ref{lem:Rparity_proof}) can be thought of as a product of
$\sigma$-derivatives multiplied by the $\sigma$-function itself
raised to a power.

If $n$ is odd then $m$ must be also, making the denominator of
(\ref{lem:Rparity_proof}) odd.  Each term in the numerator must
contain an even number of $\sigma_{\rm e}$-derivatives, paired with an odd
number of $\sigma_{\rm o}$-derivatives.  Hence, since $m$ is odd, any
$\sigma$-functions in the term must be raised to an even power.  So
overall each term in the numerator is even and hence
$\mathcal{R}^{[m]}_{i_1,\dots,i_n}$ is odd overall.

If $n$ is even then we must consider the two choices for $m$
separately. If $m$ is even then so is the denominator of
(\ref{lem:Rparity_proof}).  Each term in the numerator must have an
even number of $\sigma$-derivatives overall.  If there are an odd
number of $\sigma_{\rm e}$-derivatives then there are an even number of
$\sigma_{\rm o}$-derivatives and hence, since $m$ is even, a
$\sigma$-function raised to an odd power.  Then each term in the
numerator is even overall.  If there were an even number of
$\sigma_{\rm e}$-derivatives then there are an odd number of
$\sigma_{\rm o}$-derivatives and any remaining $\sigma$-function is raised
to an even power.  Either way, each term in the numerator is even
overall and hence so is $\mathcal{R}^{[m]}_{i_1,\dots,i_n}$.
Finally, if $m$ were odd then the same analysis will this time show that each term in the numerator is odd.  However, the denominator will also be odd in this case and so $\mathcal{R}^{[m]}_{i_1,\dots,i_n}$ is still even overall.
\end{proof}

\subsection*{Acknowledgments}

We acknowledge Mr. Lachlan Walker who contributed some preliminary
work to proving that $Q$-functions are Abelian.  In particular he
proved equation (\ref{eq:SigPerLem}), and derived a formula for the
8-index $Q$-functions in terms of $\wp$-functions which helped
motivate equation (\ref{eq:Q_to_P}).  We would also like to thank the two anonymous referees for their useful comments.

\pdfbookmark[1]{References}{ref}
\LastPageEnding

\end{document}